\documentclass[smallcondensed]{svjour3}
\usepackage[english]{babel}
\usepackage{amsmath}
\usepackage{graphicx}
\usepackage{amssymb}
\journalname{Celest Mech Dyn Astron}

\begin{document}

\title{A comparison of different indicators of chaos based on the deviation vectors. Application to symplectic mappings}

\titlerunning{A comparison of different indicators of chaos based on the deviation vectors. Application to symplectic mappings.}

\author{N. P. Maffione \and L. A. Darriba \and P. M. Cincotta \and C. M. Giordano}

\authorrunning{N.P. Maffione}

\institute{N. P. Maffione \at
Grupo de Caos en Sistemas Hamiltonianos. Facultad de Ciencias Astron\'omicas y Geof\'{\i}sicas. \\
Universidad Nacional de La Plata. \\
Instituto de Astrof\'{\i}sica La Plata (UNLP, CONICET--CCT La Plata) \\
\email{nmaffione@fcaglp.unlp.edu.ar}
\and L. A. Darriba \at
Grupo de Caos en Sistemas Hamiltonianos. Facultad de Ciencias Astron\'omicas y Geof\'{\i}sicas. \\
Universidad Nacional de La Plata. \\
Instituto de Astrof\'{\i}sica La Plata (UNLP, CONICET--CCT La Plata) \\
\email{ldarriba@fcaglp.unlp.edu.ar}
\and P. M. Cincotta \at 
Grupo de Caos en Sistemas Hamiltonianos. Facultad de Ciencias Astron\'omicas y Geof\'{\i}sicas. \\
Universidad Nacional de La Plata. \\
Instituto de Astrof\'{\i}sica La Plata (UNLP, CONICET--CCT La Plata)\\
\email{pmc@fcaglp.unlp.edu.ar}
\and C. M. Giordano \at 
Grupo de Caos en Sistemas Hamiltonianos. Facultad de Ciencias Astron\'omicas y Geof\'{\i}sicas. \\ 
Universidad Nacional de La Plata. \\
Instituto de Astrof\'{\i}sica La Plata (UNLP, CONICET--CCT La Plata) \\
\email{giordano@fcaglp.unlp.edu.ar}}
\date{\textbf{Celestial Mechanics \& Dynamical Astronomy, in press} ("The final publication will be soon available at \textit{springerlink.com}")}
\maketitle

\abstract{The aim of this research work is to compare the reliability of several variational indicators of chaos on mappings. The Lyapunov Indicator (LI); the Mean Exponential Growth factor of Nearby Orbits (MEGNO); the Smaller Alignment Index (SALI); the Fast Lyapunov Indicator (FLI); the Dynamical Spectra of stretching numbers (SSN) and the corresponding Spectral Distance (\textit{D}); and the Relative Lyapunov Indicator (RLI), which is based on the evolution of the difference between two close orbits, have been included.

The experiments presented herein allow us to reliably suggest a group of chaos indicators to analyze a general mapping. We show that a package composed of the FLI and the RLI (to analyze the phase portrait globally) and the MEGNO and the SALI (to analyze orbits individually) is good enough to make a description of the systems' dynamics.}


\section{Introduction}\label{introduction}
A key element in the analysis of the behavior of a given dynamical system lies in the possibility to accurately determine the regular or chaotic nature of its trajectories. However, such characterization often proves to be a rather subtle and complicated problem. Therefore, any technique that allows us to locate regions where chaotic motion is probable is very useful. Several of such techniques (chaos indicators, hereafter CIs) have been developed during the last few years.

Since the early work of H\'enon and Heiles (1964), the development of CIs has grown exponentially. In the case of two degrees of freedom, the basic tool is the graphical treatment through the Poincar\'e Surfaces of Section. This approach has been extended to systems with three degrees of freedom (Froeschl\'e 1970a; Froeschl\'e 1972). It is known that it has severe restrictions when dealing with systems with more degrees of freedom, though. Therefore, it is important to have at hand techniques that do work irrespective of the dimension of the problem.

The early non graphical methods for the detection of chaos were based on the concept of exponential divergence. Then, the introduction of the Lyapunov Characteristic Exponents (LCEs) and their numerical implementation (Benettin et al. 1980; Skokos 2010) was one of the first major contributions towards the search for chaos. In other words, the understanding of the dynamical behavior of some regions of the phase space relies, e.g., on the computation of at least the largest Lyapunov Characteristic Exponent (lLCE) for a large number of orbits. This might be a time-consuming process. Moreover, we can only reach a truncated value for the lLCE, because the integration time is finite. Therefore, it is interesting to define new algorithms at least as reliable as the truncated version of the lLCE (i.e. Lyapunov Indicators, hereafter LIs, Benettin et al. 1976), but cheaper in computational time. Many new techniques are now available: the Mean Exponential Growth factor of Nearby Orbits (MEGNO) developed by Cincotta and Sim\'o (2000)\footnote{See also Cincotta et al. (2003); Giordano and Cincotta (2004); Go\'zdziewski et al. (2005); Gayon and Bois (2008); Lema\^itre et al. (2009); Hinse et al. (2010); Maffione et al. (2011).}; the Smaller Alignment Index (SALI) by Skokos (2001)\footnote{See also Skokos et al. (2007); Sz\'ell et al. (2004); Bountis and Skokos (2006); Carpintero (2008); Antonopoulos et al. (2010).}; the Fast Lyapunov Indicator (FLI) introduced by Froeschl\'e et al. (1997)\footnote{See also Guzzo et al. (2002); Froeschl\'e et al. (2006); Paleari et al. (2008); Todorovi\'c et al. (2008); Lega et al. (2010).}; the Spectral Distance (\textit{D}) by Voglis et al. (1999) and the Spectra of Stretching Numbers (SSN)\footnote{See Voglis and Contopoulos (1994); Contopoulos and Voglis (1996); Contopoulos and Voglis (1997); Voglis et al. (1998).} or sometimes called Local Lyapunov Characteristic Numbers (LLCN)\footnote{See Froeschl\'e et al. (1993); Froeschl\'e et al. (2006); Todorovi\'c et al. (2008).}. Finally, we include the Relative Lyapunov Indicator (RLI, S\'andor et al. 2000; Sz\'ell et al. 2004; S\'andor et al. 2004; S\'andor et al. 2007), which is not a variational indicator like the others, but it is based on the evolution of two different but very close orbits. 

These techniques are just a sample of all the indicators used in the study of dynamical systems. The introduction of spectral indicators, such as the Frequency Map Analysis (FMA) by Laskar (e.g. Laskar 1990; Papaphilippou and Laskar 1996; Papaphilippou and Laskar 1998), and updates of the above techniques will be addressed in future reasearch works. Such updates will include the Generalized Alignment Index (GALI, Skokos et al. 2007) as a generalization of the SALI; the Orthogonal Fast Lyapunov indicator (OFLI, Fouchard et al. 2002) and the OFLI$^2_{TT}$ (Barrio 2005), which are improvements of the FLI. The Average Power Law Exponent (APLE, Lukes-Gerakopoulos et al. 2008) will also be considered. 

The analysis is performed over two different 4D mappings: a variant of Froeschl\'e's symplectic mapping, hereafter vFSM (Froeschl\'e 1972; Contopoulos and Giorgilli 1988; Skokos et al. 1997; Skokos 2001), and a system comprising two coupled standard mappings. 

The work is organized as follows: the reliability of the thresholds is studied in Section \ref{Thresholds study}. We compute the number of chaotic orbits given by every CI with their concomitant threshold and examine the variation of the chaotic component by means of a small change in the critical value. 

Once the analysis of the thresholds is done, the speed of convergence and the resolving power of the techniques can be accurately determined. Therefore, in Section \ref{Qualitative study-representative groups}, we use the information gathered from the time evolution curves to compare the performances of the indicators based on these two fundamental characteristics. The small size of the sample allows for this kind of analysis. 

In Section \ref{Qualitative study-statistical sample}, those characteristics are further analyzed by considering the information coming only from the final values of the techniques. Thus, big samples of orbits can be studied (the one selected for this work consists of $10^6$ orbits). The previous experiments of Section \ref{Qualitative study-representative groups} show the performances of the CIs with orbits that behaved well. 

We are also concerned with the study of the CIs under very complex scenarios. Hence, in Section \ref{Extreme conditions} we study the techniques over two very complex schemes: inside a stochastic layer in the main resonance and in a region populated with sticky chaotic orbits. 

Finally, in Section \ref{A case of mild chaos} we study the dependency of the \textit{D} and the RLI on their free-parameters. They are the only techniques from the package that need a previous user-choice procedure to configure the algorithms for their computation. We discuss the results in Section \ref{Discussion}. 

Although we have applied the CIs to two 4D dimensional mappings, every technique considered herein could be applied to any ND mapping but the \textit{D}. The \textit{D} is not suitable for 2D mappings (see, e.g., Skokos 2001).

\section{The reliability on the thresholds}\label{Thresholds study}
The speed of convergence, the sensitivity to hyperbolicity and stability levels (or resolving power), a reliable threshold and the computing time might be the most important characteristics that make a CI suitable for a given study.

We will deal with the CIs' thresholds first, because they are fundamental to study properly the other main characteristics. The following study is accomplished on the vFSM adopting a sample of $10^6$ initial conditions.  

The vFSM is defined by the following equations (mod$2\pi$):

$$x'_1=x_1+x_2$$
$$x'_2=x_2-\nu\cdot\sin(x_1+x_2)-\mu\cdot[1-\cos(x_1+x_2+x_3+x_4)]$$
$$x'_3=x_3+x_4$$
$$x'_4=x_4-\kappa\cdot\sin(x_3+x_4)-\mu\cdot[1-\cos(x_1+x_2+x_3+x_4)].$$

The parameters used for the mapping are $\nu=0.5$, $\kappa=0.1$ and $\mu=10^{-3}$ (see Skokos 2001 for further details).

We use the following configuration for the computation of the experiments unless stated otherwise: four numbers of iterations, i.e., $10^3$, $5\times 10^3$, $10^4$ and $10^5$ iterations. The initial separation taken for the calculation of the RLI is $10^{-12}$ (see S\'andor et al. 2004). The \textit{D} is computed over intervals of 100 iterations and the number of cells considered for the generation of the histograms for the SSN is $10^3$. The initial deviation vectors are: (1,1,1,1), (1,0,0,0), (0,1,0,0), (0,0,1,0). Froeschl\'e et al. 1997 and Froeschl\'e et al. 1997a showed that the first part of the computation of the lLCE is enough to discriminate between chaotic and regular orbits, and introduced the FLI. However, as Froeschl\'e et al. stated, on using such techniques during short times, some dependency on the initial conditions of the deviation vectors may be found. Therefore, it is important to keep the same initial deviation vectors for the whole sample along the individual experiments (Froeschl\'e and Lega 2000). The version of the MEGNO here considered is the MEGNO(2,0) (see Cincotta et al. 2003).

We applied the time-dependent threshold of the LI (Table \ref{tableind}) with $N$ being the number of iterations. It is known that an empirical adjustment of the LI's theoretical threshold is strongly advisable. Yet, our choice here is to avoid this fine tunning and to consider the raw theoretical estimation for the sake of a fair comparison. The critical value used for the RLI (Table \ref{tableind}) was computed following S\'andor et al. 2004 and the remarks discussed herein in Section \ref{A case of mild chaos}. The way to determine the time-dependent threshold for the \textit{D} is not known yet. In the case of the MEGNO(2,0), the threshold is a fixed value (Table \ref{tableind}) which also needs an empirical adjustment. However, as we did for the LI, we will use the theoretical fixed value. For the SALI there are two thresholds commonly used, namely, $10^{-12}$ and $10^{-4}$ (see e.g. Skokos et al. 2004). In between the orbits are called ``sticky chaotic''. Nevertheless, we will consider them also as chaotic orbits. Then, the threshold to analyze is $10^{-4}$ which separates regular from chaotic and sticky chaotic orbits. Once again, this is our choice to avoid any advantages in taking more than one critical value. The threshold associated with the FLI (sometimes used with two thresholds also, see Paleari et al. 2008) is time-dependent and has the formulae presented on Table \ref{tableind} (hereafter in the computation of the FLI, we do not consider the logarithm which is usually used in its definition, see e.g. Todorovi\'c et al. 2008).

\begin{table}[!ht]\centering
\begin{tabular}{cc}
\hline\hline  \vspace*{-2ex} \\ 
\emph{} CI  & Threshold \vspace*{1ex} \\ 
\hline 
 LI & $ln(N)/N$ \vspace*{1ex} \\
\hline 
 RLI & $10^{-12}$ \vspace*{1ex} \\
\hline 
 MEGNO(2,0) & $0.5$ \vspace*{1ex} \\
\hline 
 SALI & $10^{-4}$ \vspace*{1ex} \\
\hline 
 FLI & $N$ \vspace*{1ex} \\
\hline\hline  \vspace*{-4ex} 
\end{tabular}
\vspace{3mm}
\caption{Thresholds for the LI, the RLI, the MEGNO(2,0), the SALI and the FLI.}
\label{tableind}
\end{table}

\begin{table}[!ht]\centering
\begin{tabular}{cccccccc}
\hline\hline  \vspace*{-2ex} \\ 
\emph{} CI  & $N$ & Threshold & Chaos (\%) & Chaos (\%) A & Chaos (\%) B \vspace*{1ex} \\ 
\hline
 LI & $10^3$ & 6.9077555E-03 & 69.4299 & 68.7814 & 70.0918 \vspace*{1ex} \\
\hline 
  & $5\times 10^3$ & 1.7034386E-03 & 73.2632 & 72.7674 & 73.8041 \vspace*{1ex} \\
\hline 
  & $10^4$ & 9.2103402E-04 & 73.5061 & 72.9797 & 74.0848 \vspace*{1ex} \\
\hline 
  & $10^5$ & 1.1512925E-04 & 73.8847 & 73.2543 & 74.5822 \vspace*{1ex} \\
\hline 
 RLI & $10^3$ & $10^{-12}$ & 49.9296 & 49.7875 & 50.0667 \vspace*{1ex} \\
\hline 
  & $5\times 10^3$ & $10^{-12}$ & 71.7731 & 71.7542 & 71.7912 \vspace*{1ex} \\
\hline 
  & $10^4$ & $10^{-12}$ & 72.0777 & 72.0617 & 72.0961 \vspace*{1ex} \\ 
\hline 
  & $10^5$ & $10^{-12}$ & 72.34 & 72.3255 & 72.3549 \vspace*{1ex} \\
\hline
 MEGNO(2,0) & $10^3$ & $0.5$ & 85.9964 & 78.8211 & 87.7021 \vspace*{1ex} \\
\hline
  & $5\times 10^3$ & $0.5$ & 91.4468 & 77.0322 & 94.7615 \vspace*{1ex} \\
\hline
  & $10^4$ & $0.5$ & 91.8992 & 75.0808 & 95.8364 \vspace*{1ex} \\
\hline
  & $10^5$ & $0.5$ & 92.7217 & 71.6947 & 98.2398 \vspace*{1ex} \\
\hline
 SALI & $10^3$ & $10^{-4}$ & 14.7913 & 14.8604 & 14.7232  \vspace*{1ex} \\
\hline 
  & $5\times 10^3$ & $10^{-4}$ & 68.3665 & 68.3676 & 68.3658  \vspace*{1ex} \\
\hline 
  & $10^4$ & $10^{-4}$ & 68.8171 & 68.8187 & 68.8168  \vspace*{1ex} \\
\hline 
  & $10^5$ & $10^{-4}$ & 69.778 & 69.7786 & 69.7776 \vspace*{1ex} \\
\hline
 FLI & $10^3$ & $10^3$ & 69.7826 & 69.6846 & 69.8806  \vspace*{1ex} \\
\hline 
  & $5\times 10^3$ & $5\times 10^3$ & 73.3382 & 73.2752 & 73.4007  \vspace*{1ex} \\
\hline 
  & $10^4$ & $10^4$ & 73.571 & 73.5102 & 73.6329  \vspace*{1ex} \\
\hline 
  & $10^5$ & $10^5$ & 73.9327 & 73.8734 & 73.9905 \vspace*{1ex} \\
\hline\hline  \vspace*{-4ex} 
\end{tabular}
\vspace{3mm}
\caption{Evolution in time of the percental variation of the chaotic component for three different values of the associated thresholds (see text for details).}
\label{tableporcentajes}
\end{table}

In Table \ref{tableporcentajes} we show the CIs' values for the last iteration (or final values) for the array of $10^6$ initial conditions and four $N$s: $10^3$, $5\times 10^3$, $10^4$ and $10^5$ iterations. There, we resume the time evolution of the percental variation of the chaotic component for three different values of the associated threshold. The third column displays the value of the threshold whereas the fourth column of Table \ref{tableporcentajes} (``Chaos (\%)'') corresponds to the percentage of chaotic orbits which use the raw estimate of the (theoretical or empirical) threshold. The last two columns are the percentages of the chaotic component according to a change in the threshold by $\pm1\%$ (a change by $+1\%$ is labeled A and a change by $-1\%$ is labeled B on last column), emulating the fine tunning of the critical value. 

It is interesting to notice that after $5\times 10^3$ iterations and despite the adjustments of the thresholds, all the indicators but the MEGNO(2,0) mostly agree in the percental variation of the chaotic orbits (last three columns of Table \ref{tableporcentajes}), i.e. between $68\%$ and $75\%$. The MEGNO(2,0) shows a very high fraction of chaotic orbits either if we adopt the theoretical fixed threshold (fourth column) or the modified threshold of the last column (B). On the other hand, if we use a threshold closer to $0.505$ (fifth column, A), the MEGNO(2,0) shows a better approach to the prevailing percentage of chaos. Both facts tell us that not only the theoretical fixed threshold of the MEGNO(2,0) needs to be carefully adjusted, but also that its threshold is the most sensitive to an experimental rearrangement.

\begin{figure}[ht!]
\begin{center}
\begin{tabular}{c}
\hspace{-5mm}\resizebox{90mm}{!}{\includegraphics{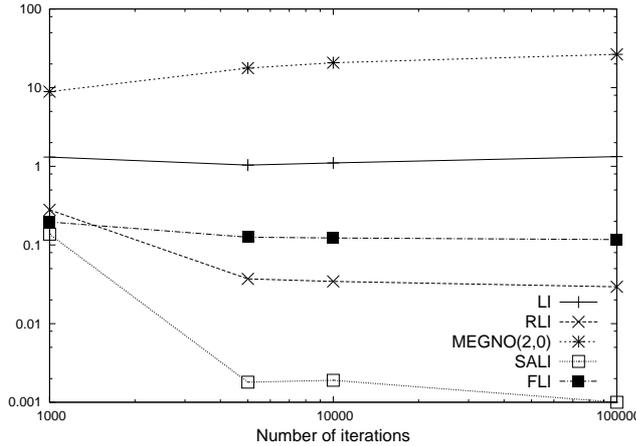}}
\end{tabular}
\caption{Time evolution of the difference between the percentage of chaotic orbits using an adjusted threshold by $+1\%$ and an adjusted threshold by $-1\%$.} 
\label{robustezvc}
\end{center}
\end{figure}

In Fig. \ref{robustezvc} we show the time evolution of the difference between the percentages of chaotic orbits found in A and the percentages of chaotic orbits found in B (fifth and last columns of Table \ref{tableporcentajes}). Then, the weakness of the theoretical fixed threshold for the MEGNO(2,0) becomes evident (curve with the highest values). This is a consequence of the asymptotic nature of the threshold. It becomes clear that the theoretical threshold of the MEGNO(2,0) is just an estimation of the value to be used. On the other hand, we find that the RLI and the SALI have very reliable thresholds, despite their empirical nature (curves with the lowest values on Fig. \ref{robustezvc}). The SALI has the robustest threshold according to Fig. \ref{robustezvc}. However, the media of the percentages for the chaotic component (see Table \ref{tableporcentajes}, last three columns) is below the medias given by the LI, the RLI or the FLI if the $N$ is above $5\times 10^3$. Thus, among the latter CIs, the RLI shows the most reliable threshold in the experiment. 

\section{Study of some representative orbits through the CIs' time evolution}\label{Qualitative study-representative groups}
In this section, we study other aspects of a CI: the speed of convergence and the sensitivity to distinguish different kinds of motions. We use the time evolution curves because of the reduced size of the sample selected. This first sample consists of ten orbits, five of them are chaotic orbits and five are regular orbits according to a convergent LI (the $N$ chosen to guarantee the convergence of the LI is $10^{5}$ iterations which is much longer than the actual convergent time). 

On a $(x_1,x_2)$ space, the projections of every chaotic orbit of the sample fulfill the connected chaotic component (see e.g. Fig. 3 in Skokos 2001); they are not ergodic, though. The chaotic orbit with initial condition $x_2=-0.5$ is the only one with an amount of stickiness (see e.g. Kov\'acs and \'Erdi 2009; Contopoulos and Harsoula 2010) around the main stability island. In the case of the regular component, we have invariant curves. An examination of the plots on a $(x_1,x_2)$ space, shows that the regular orbit with initial condition $x_1=2$ is very closed to a stochastic layer and therefore presents some characteristics of a stochastic orbit. The chaotic orbit with initial condition $x_2=-0.5$ and the regular orbit with initial condition $x_1=2$ are selected for further study. We present their initial conditions in Table \ref{tableinicon}. 

\begin{table}[!ht]\centering
\begin{tabular}{ccccc}
\hline\hline\vspace*{-2ex} \\ 
\emph{} Nature of the orbit & $x_1$ & $x_2$ & $x_3$ & $x_4$ \vspace*{1ex} \\ 
\hline
 Chaotic orbit & 3 & -0.5 & 0.5 & 0 \vspace*{1ex} \\
\hline
 Regular orbit & 2 & 0 & 0.5 & 0 \vspace*{1ex} \\
\hline\hline  \vspace*{-4ex} 
\end{tabular}
\vspace{3mm}
\caption{Initial conditions for the chaotic orbit and the regular orbit.}
\label{tableinicon}
\end{table}

\begin{figure}[ht!]
\begin{center}
\begin{tabular}{cc}
\hspace{-5mm}\resizebox{63mm}{!}{\includegraphics{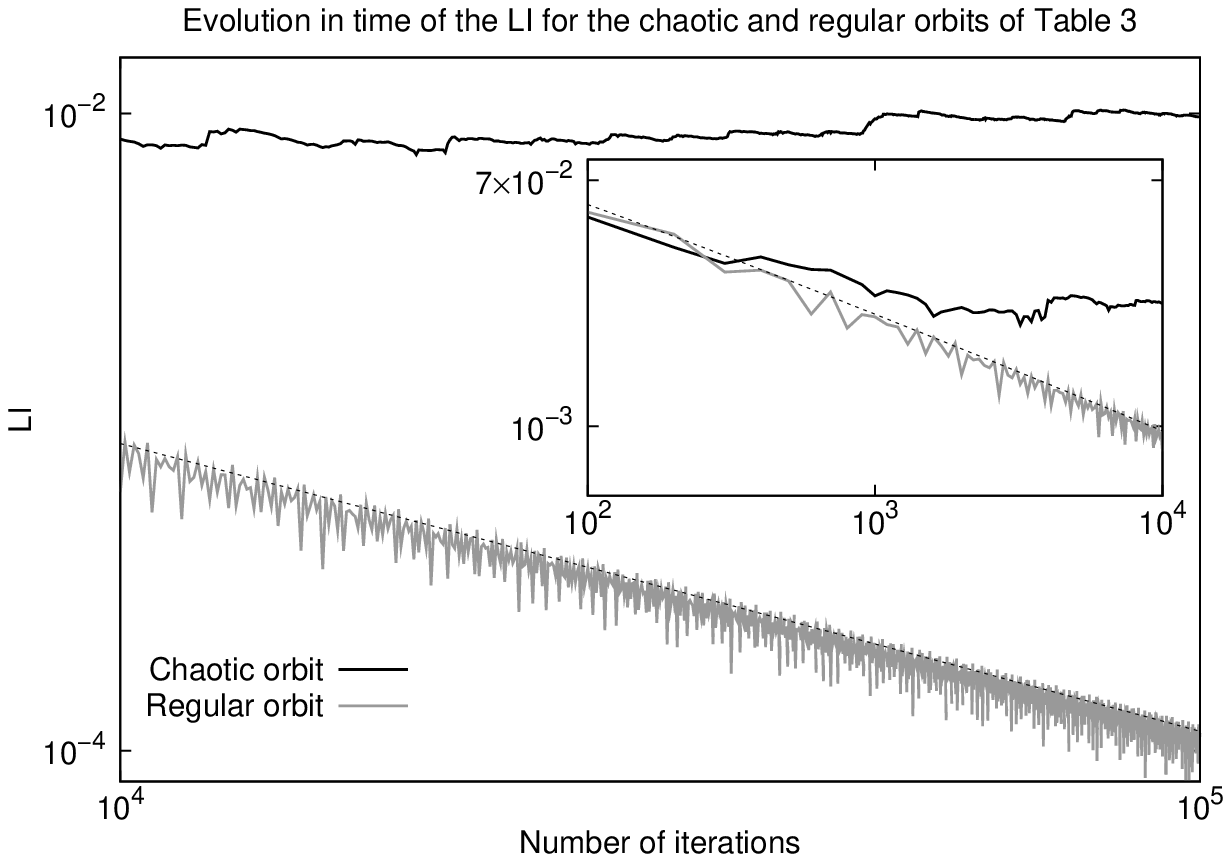}}& 
\hspace{-5mm}\resizebox{63mm}{!}{\includegraphics{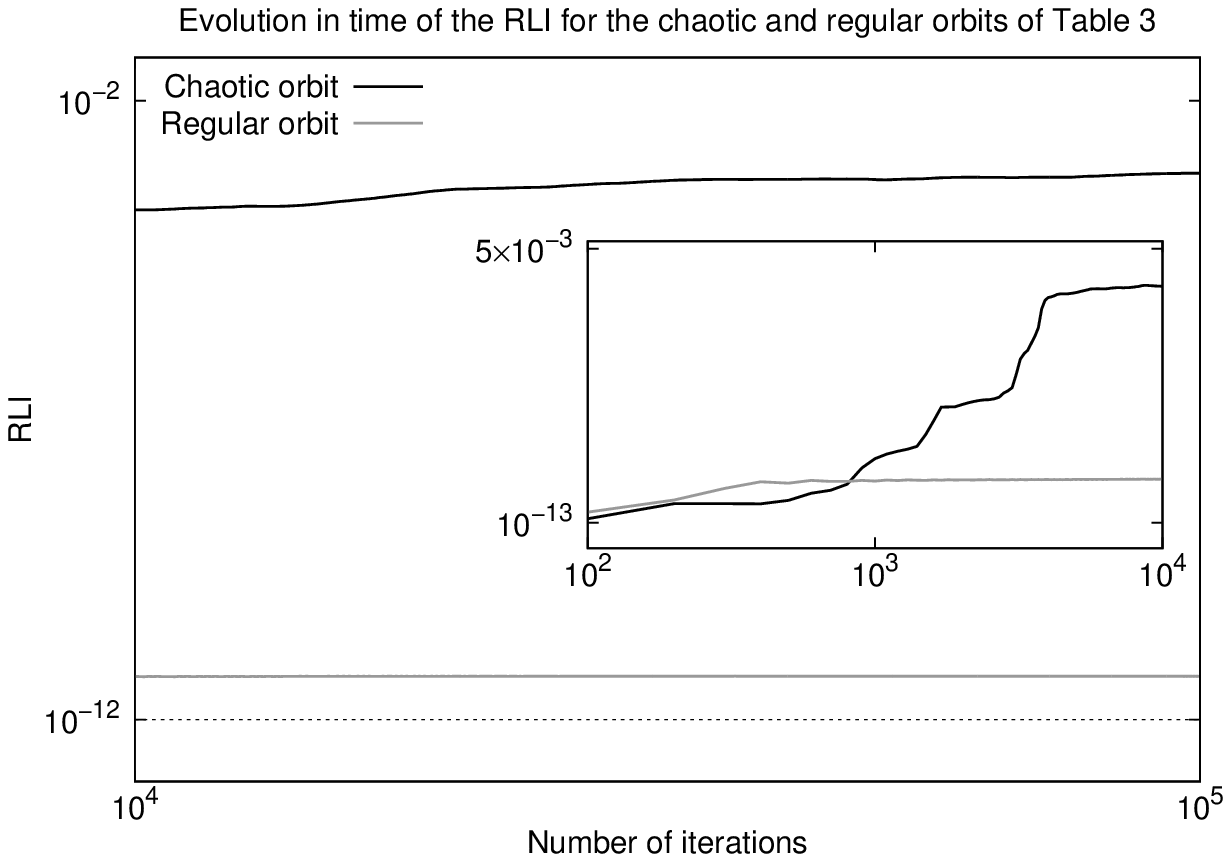}}\\
\hspace{-5mm}\resizebox{63mm}{!}{\includegraphics{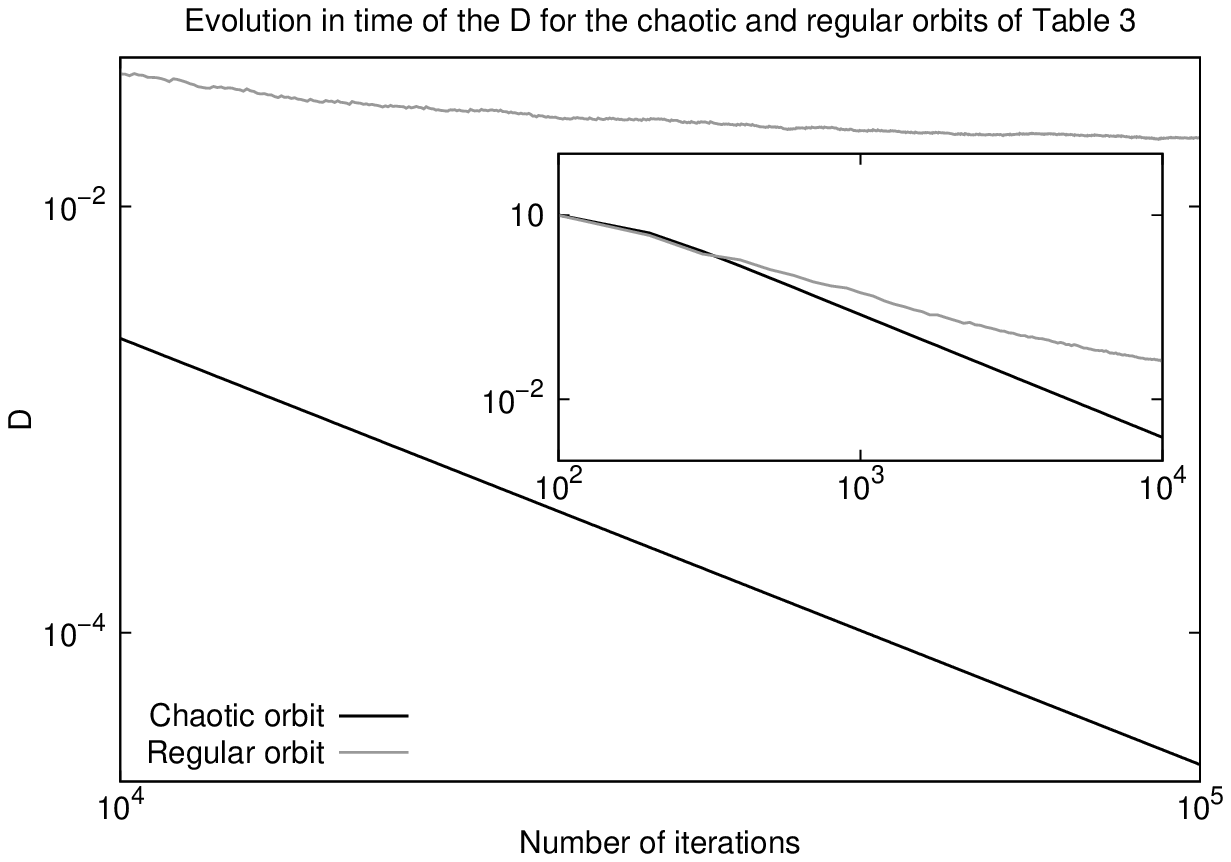}}& 
\hspace{-5mm}\resizebox{63mm}{!}{\includegraphics{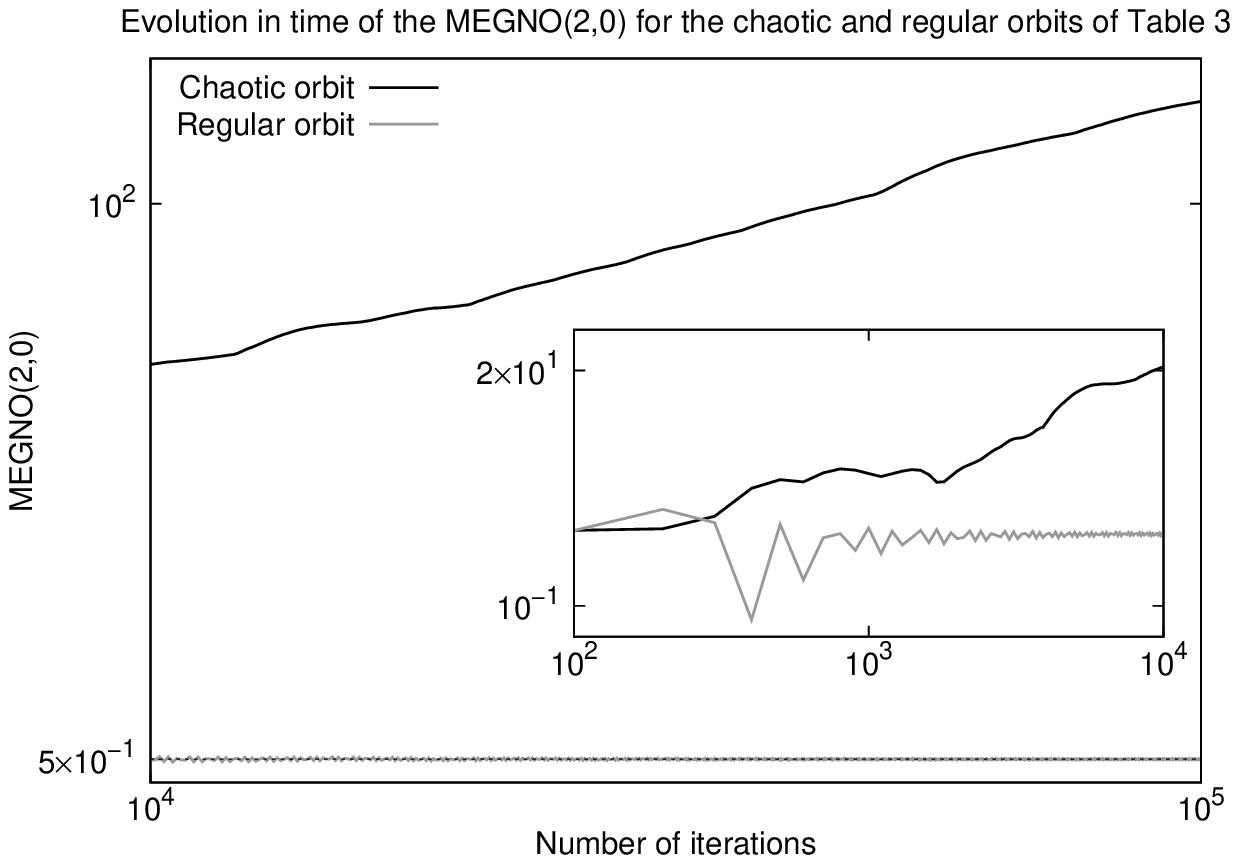}}\\
\hspace{-5mm}\resizebox{63mm}{!}{\includegraphics{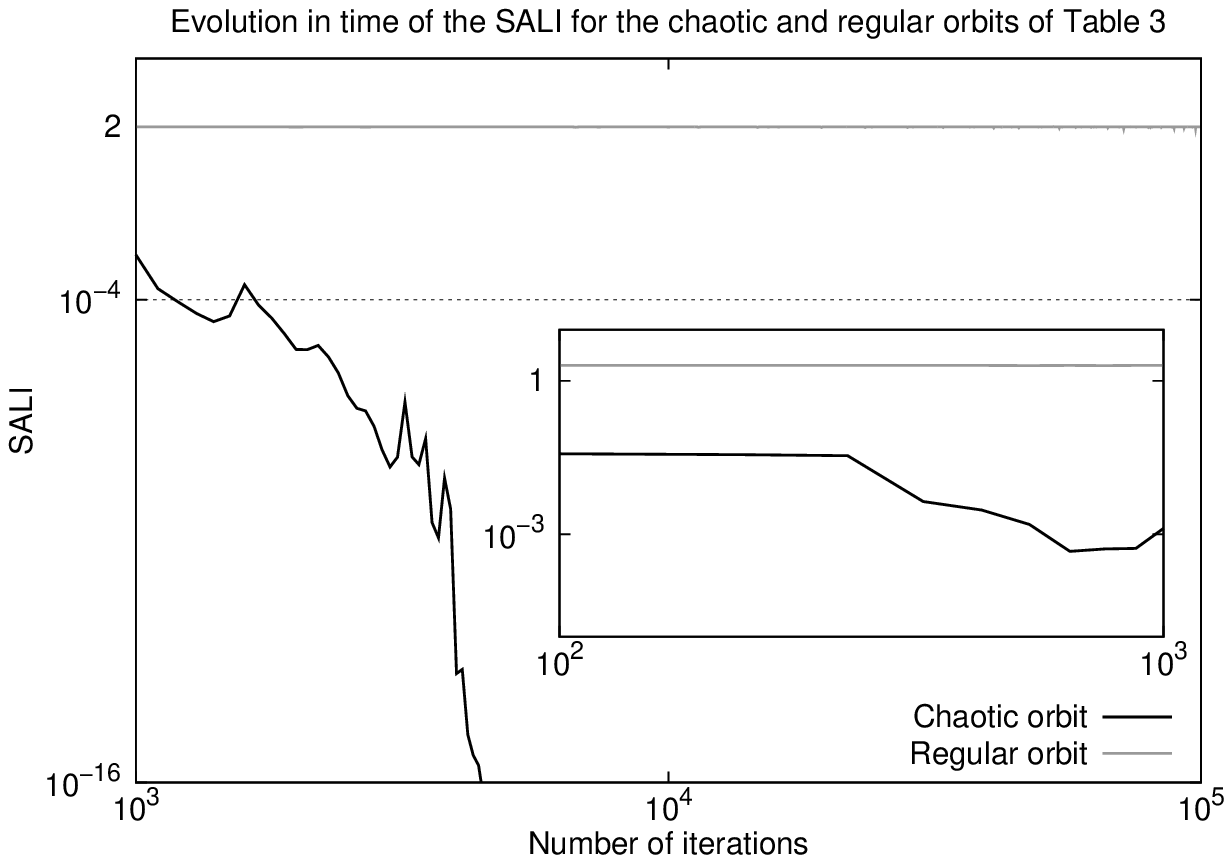}}& 
\hspace{-5mm}\resizebox{63mm}{!}{\includegraphics{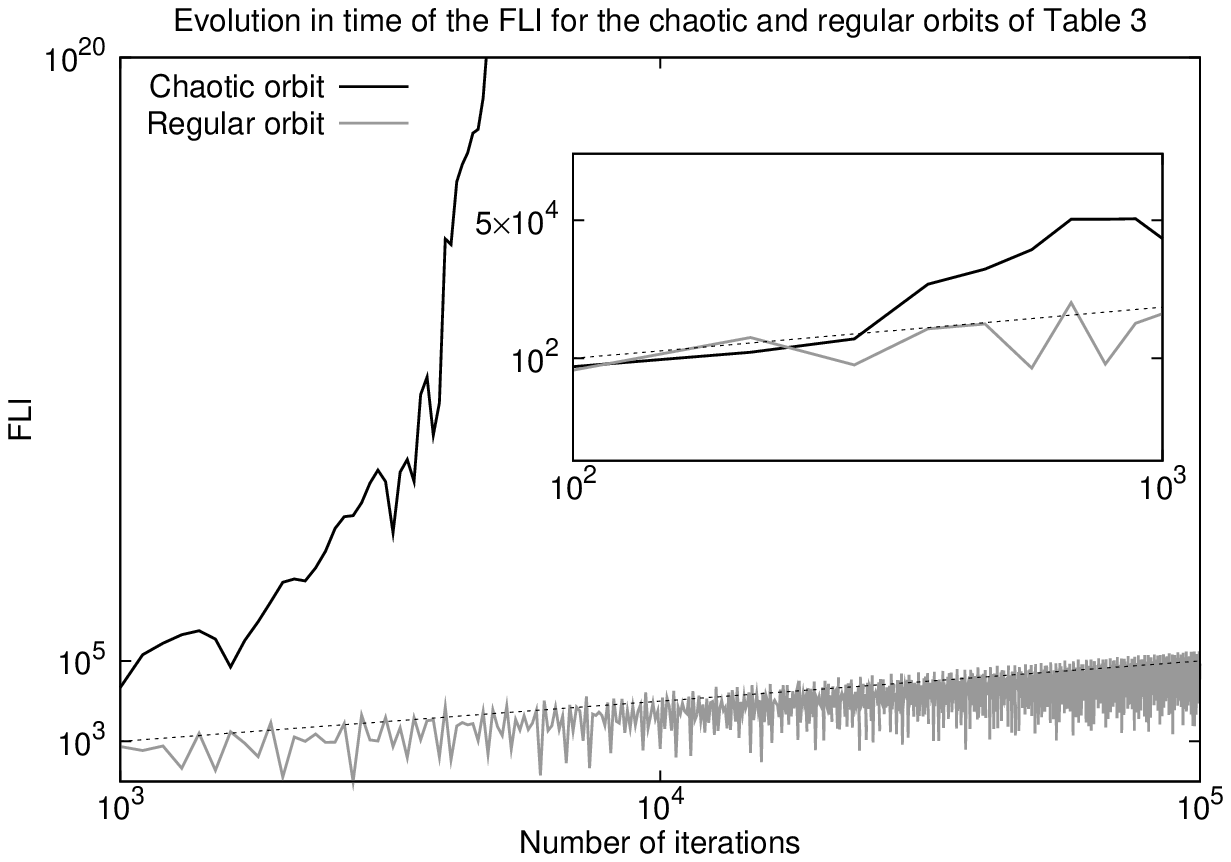}}
\end{tabular}
\caption{Time evolution of the CIs for chaotic and regular orbits on the vFSM. The thresholds are also shown like dashed curves. From top left to bottom right panels we present the LI, the RLI, the \textit{D}, the MEGNO(2,0), the SALI and the FLI, respectively. Note the different X-axis scales used for the two bottom panels.}
\label{cao.reg.cvalue}
\end{center}
\end{figure}

In Fig. \ref{cao.reg.cvalue} we plot the time evolution of the LI, the RLI, the \textit{D}, the MEGNO(2,0), the SALI and the FLI techniques for the selected orbits. The chaotic orbit is depicted in solid black and the regular orbit, in solid gray color. We also use dashed curves to plot the thresholds adopted to separate chaotic from regular motion. 

Regarding the speed of convergence, the LI characterizes the chaoticity of the orbit before $10^3$ iterations according to its theoretical threshold (i.e. $ln(N)/N$), but the constancy of the LI is only evident in some iterations after $10^3$. In the case of the regular orbit, if we follow the time-dependent threshold, the determination of its nature is made from the beginning (see top left panel of Fig. \ref{cao.reg.cvalue}). However, the separation from the chaotic orbit is made later by the $N$ above-mentioned.

The RLI increases its value for the chaotic orbit above the threshold of $10^{-12}$ from the beginning and freezes itself around $10^{-4}$ close to $10^4$ iterations. Thus, its performance is similar to that of the LI. The proximity of the regular orbit to a stochastic layer turns out to be a problem in the classification of the RLI and needs to be carefully examined. According to the RLI, this orbit, which is classified by a convergent LI as regular, should be labeled as chaotic. Since it oscillates above $10^{-12}$ (around the value $5\times 10^{-12}$, see top right panel of Fig. \ref{cao.reg.cvalue}). Nevertheless, this value is very different from those of the chaotic orbits (the five chaotic orbits used in the initial sample have convergent values between $10^{-2}$ and $10^{-4}$). Therefore, the orbits with low values but above the threshold should be evaluated separately due to a probable level of instability in their dynamics. 

The \textit{D} does not have a determined threshold, like the other CIs, to establish a more precise estimation of the speed of convergence. Moreover, the behavior of both kinds of motions is very similar in the transitory regime because the \textit{D} shows a decrease in its values independently of the nature of the orbit (with different time rates, though). As the tendency of the chaotic orbit to decrease appears before $10^3$ iterations, the speed of convergence for this CI seems to be greater than that obtained by the LI and the RLI. However, considering the similar behavior and the lack of a certain threshold, it is advisable to look at the behavior of the regular orbit. We find that the identification of its nature is made at the end of the interval ($\sim10^5$) when an almost constant value is reached, i.e. much later than the other CIs (middle left panel of Fig. \ref{cao.reg.cvalue}, see Contopoulos and Voglis 1996; Contopoulos et al. 1997; Voglis et al. 1998 for further details on the SSN, which is the basis of the \textit{D}, and Voglis et al. 1999 to review the main aspects of the \textit{D}). Ergo, the statistical basis of the \textit{D}, which leads to a later constancy in the regular orbit, and the absence of a different behavior in the transitory regime of the chaotic and regular orbits might delay some eventual classification. 

The MEGNO(2,0) tends to a theoretically deduced fixed value for the regular component, namely $0.5$, which makes it very simple to identify regular orbits and levels of stability. And it grows exponentially for the chaotic component (see Cincotta et al. 2003). This different way of identifying both kinds of motions helps to improve the speed of convergence dramatically. The chaotic orbit shows its nature around $10^3$ iterations, like the LI and the RLI, whereas the regular orbit shows an oscillatory behavior of decreasing amplitude around the threshold since the very beginning of the iterative process (see middle right panel of Fig. \ref{cao.reg.cvalue}).     

The SALI fluctuates in the interval $(0,2)$ for regular orbits, or alternatively decays exponentially at a rate related to the LCEs involved (see Skokos et al. 2007 for a complete analysis). The chaotic orbit crosses the threshold of $10^{-4}$ after $10^3$ iterations, revealing its nature. Moreover, before $10^4$ iterations the SALI reaches the computer precision and the computation is stopped. The regular orbit oscillates in the above-mentioned interval, around the value $\sim 0.05$ (bottom left panel of Fig. \ref{cao.reg.cvalue}). 

Finally, the FLI grows exponentially for chaotic motion (see e.g. Froeschl\'e et al. 1997; Froeschl\'e et al. 1997a; and Froeschl\'e and Lega 2006) and, consequently, the chaotic orbit crosses the time-dependent threshold ($N$) before $10^3$ iterations. This fact proves that its threshold is very efficient in the case of the shorter $N$s. Since a saturation number is needed to avoid overflow in the computations, we choose the value $10^{20}$, which is reached by the chaotic orbit before $10^4$ iterations (as in the SALI). The regular orbit grows with a power law instead (bottom right panel of Fig. \ref{cao.reg.cvalue}).     

Although the $N$ needed to characterize the motion (i.e. the speed of convergence) does not vary much from one indicator to the other, the resolving power deepens the differences between them. By $\sim10^5$ iterations the stickiness in the selected chaotic orbit becomes evident by the LI because of a sudden increase in its final values (see top left panel of Fig. \ref{cao.reg.cvalue}). The MEGNO(2,0) gives similar results. Such orbit can be distinguished from the rest of the chaotic orbits because it reaches the highest final value, after an incipient increase close to $10^5$ iterations. These coincident behaviors are the result of the close relation between the MEGNO and the LI (it is much faster to compute the LI through the MEGNO than the LI through the classical algorithm by Benettin et al. 1980; see e.g. Cincotta et al. 2003). Although neither the RLI and the SALI nor the FLI could distinguish this sticky behavior, the regular orbit close to a stochastic layer could be distinguished from the rest of the regular orbits of the sample by all the studied CIs. The LI, as well as the MEGNO(2,0), the SALI and the FLI, show the difference in the amplitude of the oscillations, which are bigger than in the other ordered orbits. The RLI for this regular orbit shows the highest convergent value among the regular orbits, which is more than one order of magnitude larger (notice the previous discussion of the RLI about this particular example). Finally, the \textit{D} has the lowest convergent value for this orbit among the regular orbits of the sample. However, there is no solid basis to separate it from two other regular orbits of the sample which have similar low final values and are not close to any stochastic structure.

In order to end the comparison on a small group of chaotic and regular orbits, we study their corresponding SSNs. Taking different chaotic orbits from the sample does not change the corresponding SSN. Therefore, these chaotic orbits belong to a connected domain which is not a direct result for the CIs previously considered. If we analyze the regular sample, the SSN changes for different initial conditions or even for different initial deviation vectors. This allows us to separate chaotic from regular motion (see Contopoulos and Voglis 1996; Contopoulos et al. 1997; Voglis et al. 1998; Froeschl\'e et al. 2006 and Todorovi\'c et al. 2008 for details). However, the regular orbit with initial condition $x_1=2$ has almost identical SSN profiles even when varying the initial deviation vectors. Thus it resembles a chaotic orbit because the orbit moves close to weak (and unconnected) chaotic domains inside the main stability island. Even though this unstable behavior is also observed in the aforementioned indicators, only the SSN seems to present an inaccurate result at first sight.

Finally, in this experiment in which we make use of the time evolution of the different CIs to study single orbits, the LI and the MEGNO(2,0) seem to be the most appropriate ones since they correctly identify not only the influence of the stochastic domain on the regular orbit, as the other CIs do, but also the sticky behavior of the chaotic orbit. 

\section{Study of a sample of orbits through the CIs' final values}\label{Qualitative study-statistical sample}
In the last section we studied the speed of convergence and the measurement of the hyperbolicity levels by means of the time evolution of each indicator. Such analysis provides detailed information only for a few orbits because of the common restrictions on computing times. However, as the understanding of a dynamic system is generally enhanced by the size of the studied sample, the time evolution is no longer efficient. 

In this section we only use the information provided by the CIs' final values to test their speed of convergence and resolving power. 

The sample used for this analysis consists of $10^6$ initial conditions (the same as in Section \ref{Thresholds study}). The parameters for the computation of the CIs, including the thresholds (see Table \ref{tableind}) are the same adopted in previous sections. Let us say that the \textit{D} is not considered in the first part of the current study since it lacks a well-defined threshold.

We begin by considering the speed of convergence of the CIs. In this direction, we compute the percentages of chaotic orbits of the sample for the four aforementioned $N$s. The results are presented in the fourth column of Table \ref{tableporcentajes}, in Section \ref{Thresholds study}.

We compute the differences between the certified stable percentage of chaotic orbits by $10^5$ iterations and the percentages by the other values of $N$, namely, $10^3$, $5\times 10^3$ and $10^4$. The concomitant results are shown in Table \ref{differencesporc}.

\begin{table}[!ht]\centering
\begin{tabular}{cccccc}
\hline\hline  \vspace*{-2ex} \\ 
\emph{} N  & LI (\%)& RLI (\%)& MEGNO(2,0) (\%)& SALI (\%) & FLI (\%) \vspace*{1ex} \\ 
\hline 
 $10^3$ & 4,4548 & 22,4104 & 6,7253 & 54,9867 & 4,1501 \vspace*{1ex} \\
\hline 
 $5\times 10^3$ & 0,6215 & 0,5669 & 1,2749 & 1,4115 & 0,5945 \vspace*{1ex} \\
\hline 
 $10^4$ & 0,3786 & 0,2623 & 0,8225 & 0,9609 & 0.3617 \vspace*{1ex} \\
\hline\hline  \vspace*{-4ex} 
\end{tabular}
\vspace{3mm}
\caption{Differences between the certified stable percentage of chaotic orbits by $10^5$ iterations and the percentages by the other $N$s: $10^3$, $5\times 10^3$ and $10^4$ for the several CIs.}
\label{differencesporc}
\end{table}

According to Table \ref{differencesporc}, the best approach to the final distribution of the motion by $10^3$ iterations is the one given by the FLI while the worst is that corresponding to the SALI. Since both CIs have similar behavior for chaotic orbits, the difference in the convergence must be in the threshold. The time-dependent threshold used for the FLI seems to be much more efficient than the time-independent one used for the SALI. Nevertheless, we notice that this difference shrinks very rapidly as the iteration number increases. 

After the first transient, in quasi stable regimes the RLI has the best performance due to the speed of convergence. Although the FLI starts with the highest rate of convergence, the RLI seems to be more suitable in more stable regimes. However, both the FLI and the LI follow the RLI closely. 

The remaining CIs, i.e. the MEGNO(2,0) and the SALI, have both unpaired results in the characterization of the phase portrait, as we saw in Section \ref{Thresholds study}.    

Thus, we reinforce the fact (already pointed out in Section \ref{Thresholds study}) that, for big sample of orbits, the RLI shows the most reliable speed of convergence due to its well-suited threshold. 
 
Let us now turn to the  study of the resolving power of the CI's through their final values on the sample of $10^6$ initial conditions. 

\begin{figure}[ht!]
\begin{center}
\begin{tabular}{cc}
\hspace{-5mm}\resizebox{63mm}{!}{\includegraphics{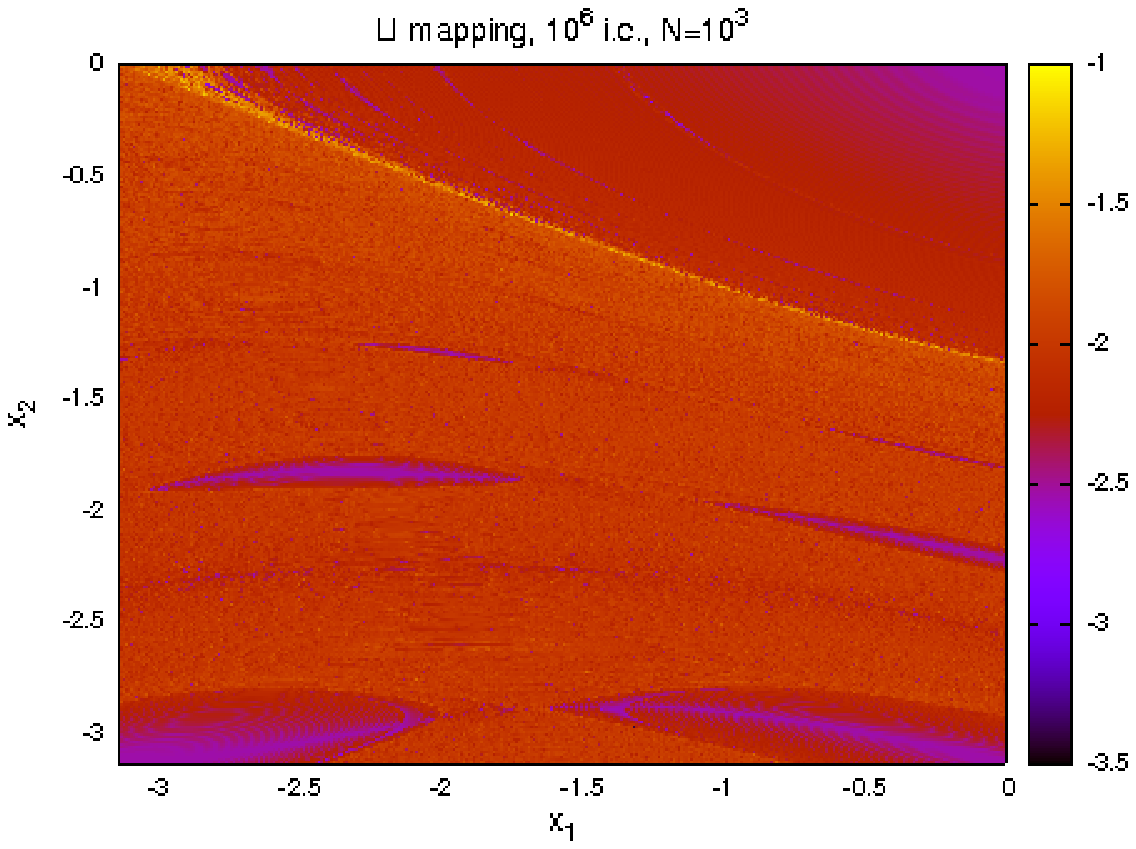}}&
\hspace{-5mm}\resizebox{63mm}{!}{\includegraphics{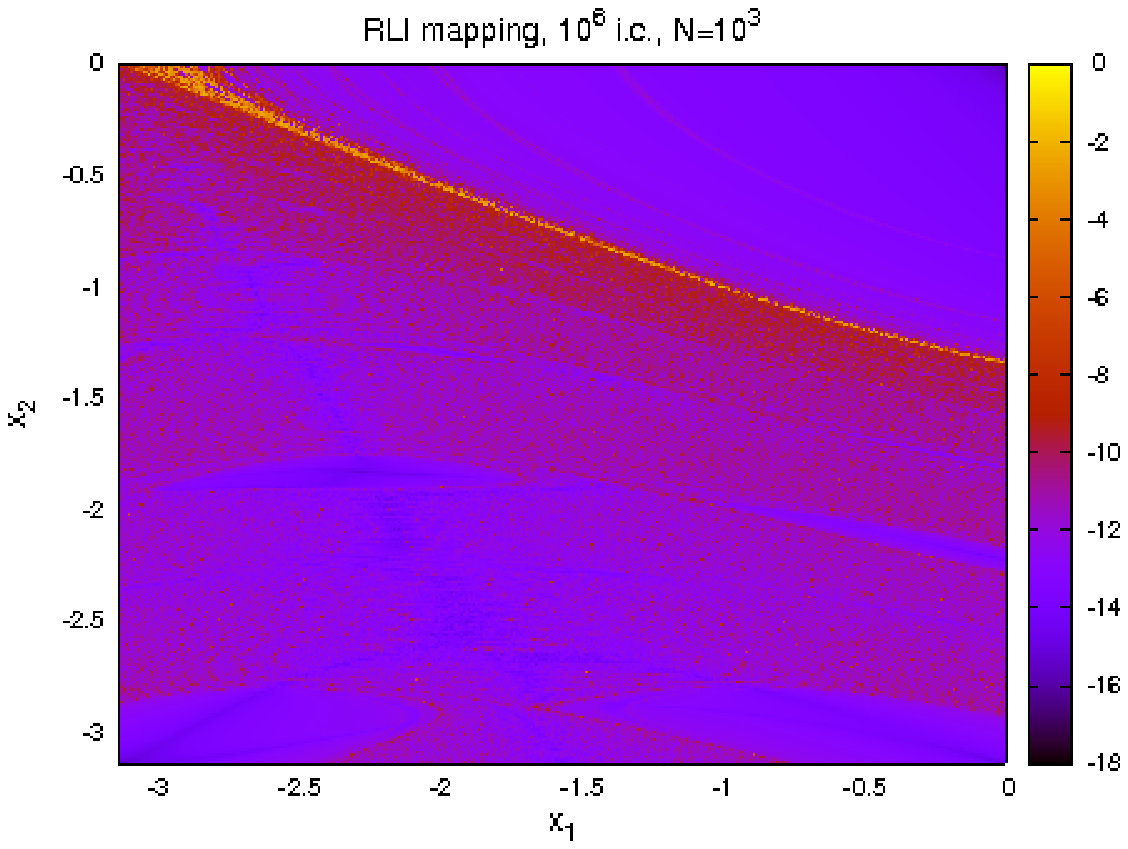}}\\ 
\hspace{-5mm}\resizebox{63mm}{!}{\includegraphics{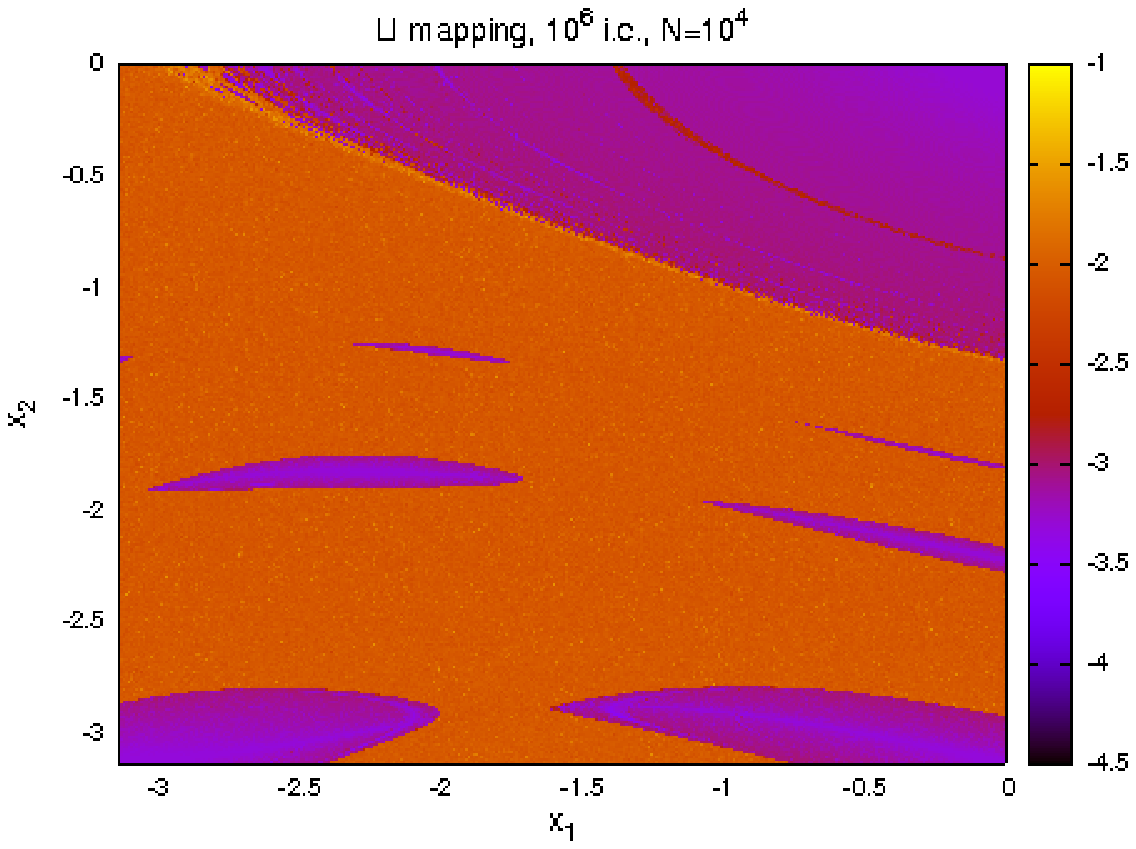}}&
\hspace{-5mm}\resizebox{63mm}{!}{\includegraphics{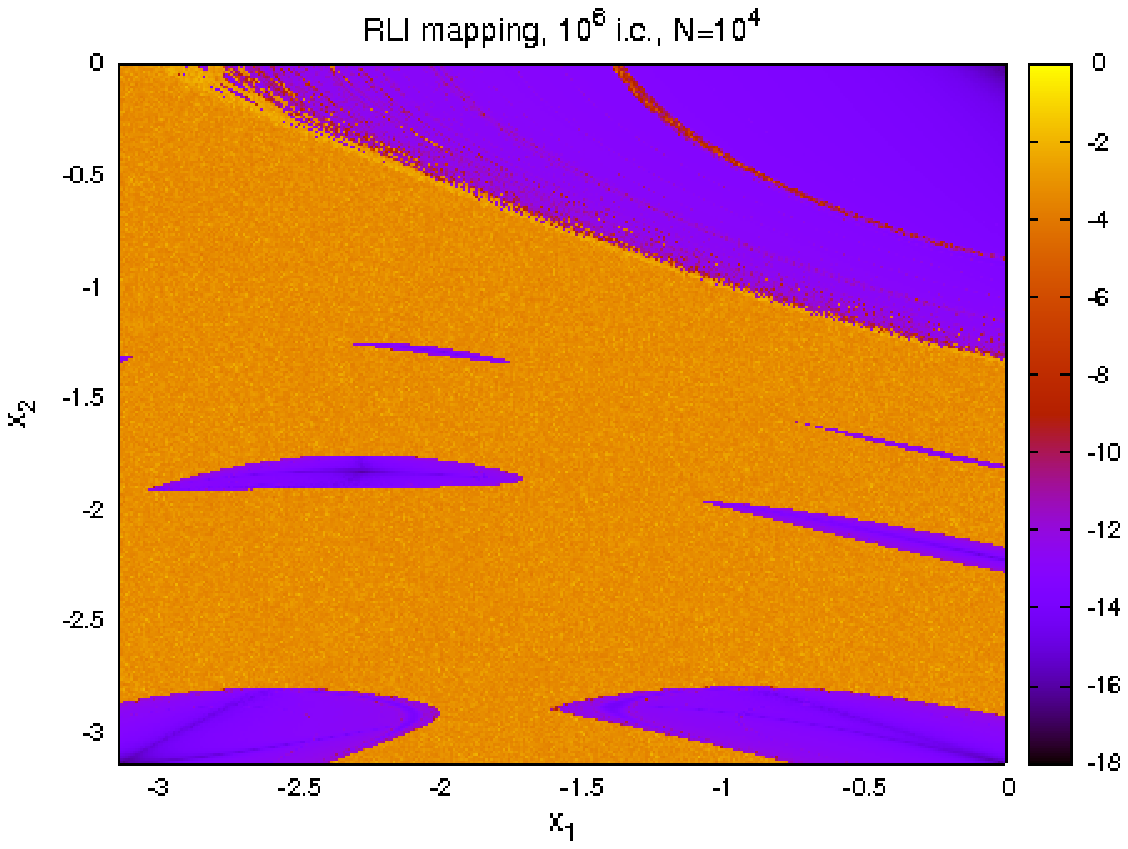}}
\end{tabular}
\caption{LI and RLI mappings on color-scale plots composed of $10^6$ initial conditions, for $10^3$ (top panels) and $10^4$ (bottom panels) iterations. On the left, the LI; on the right, the RLI, in logarithmic scale.} 
\label{liandrli1e6}
\end{center}
\end{figure}

In Fig. \ref{liandrli1e6} we present the mappings corresponding to the LI (left panels) and the RLI (right panels). These plots show snapshots at two different $N$s: $10^3$ iterations (top panels) and $10^4$ iterations (bottom panels). On the top left panel of Fig. \ref{liandrli1e6}, by $10^3$ iterations, we reach a noisy and incomplete but promising phase space portrait by the LI. Although the lack of structures like the stochastic layer inside the main stability island suggests larger $N$s, the main resonances are clearly shown. In fact, by $5\times 10^3$ iterations, not only does such stochastic layer finally appear (the concomitant figure is not included) but some structures inside the islands at the bottom are also shown. Moreover, the level of description does not change significantly by doubling the $N$ to $10^4$ iterations (bottom left panel of Fig. \ref{liandrli1e6}) which means that the LI has already reached stable values for the chaotic orbits of the sample. The main difference between both $N$s lies in the improvement of the description of the sticky chaotic orbits (e.g. in the border of the main resonance), revealing their actual chaotic nature. Finally, by $10^5$ iterations the LI does not give extra information.
 
The RLI mapping corresponding to $N=10^3$ (top right panel of Fig. \ref{liandrli1e6}) presents a very noisy phase space portrait (which is confirmed by smaller samples of $10^4$ orbits). Again, it is clear that these few iterations are not enough to separate many of the orbits in the chaotic domain from those on the regular domain (at least with the threshold selected). As regards the other number of iterations, the results do not differ very much from the results given for the LI in the previous paragraph. The level of description is similar to the one shown by the LI, and by $5\times 10^3$ iterations the phase space portrait seems to present a stable picture. Increasing the number of iterations helps to resolve the sticky chaotic orbits (bottom right panel of Fig. \ref{liandrli1e6}) but no further advantage is observed. The level of separation of the chaotic orbits belonging to the stochastic layer inside the main resonance and the regular orbits that surround them is greater than the one shown by the LI by $5\times 10^{3}$ and $10^4$ iterations, thus favoring their detection. Again, some kind of structure is seen in the high-order resonances at the bottom of the figures.

\begin{figure}[ht!]
\begin{center}
\begin{tabular}{cc}
\hspace{-5mm}\resizebox{63mm}{!}{\includegraphics{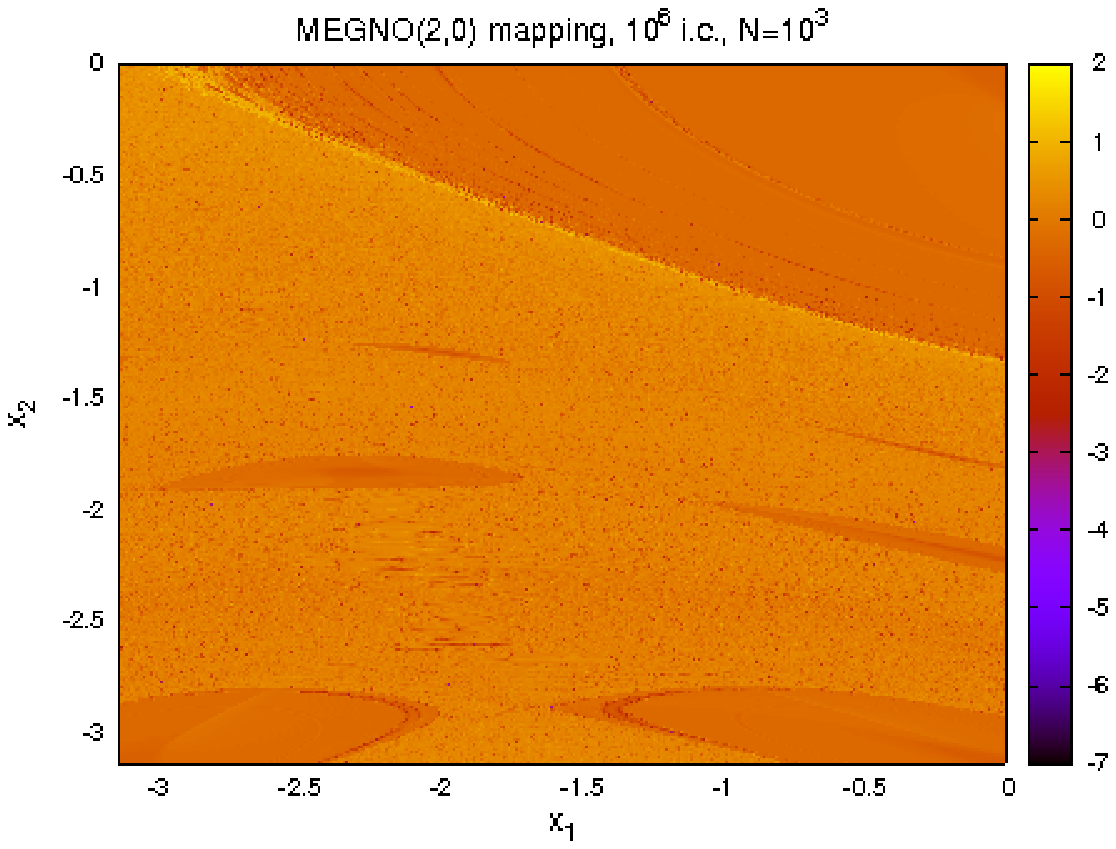}}&
\hspace{-5mm}\resizebox{63mm}{!}{\includegraphics{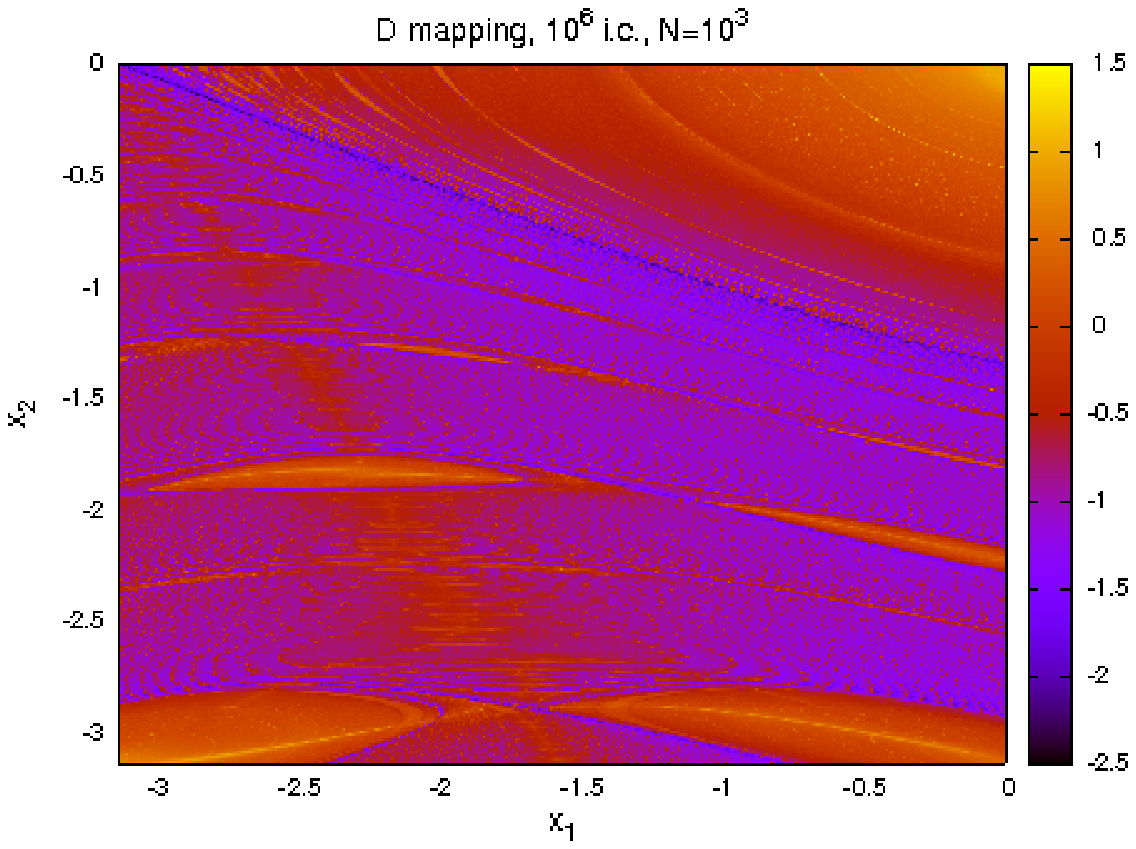}}\\ 
\hspace{-5mm}\resizebox{63mm}{!}{\includegraphics{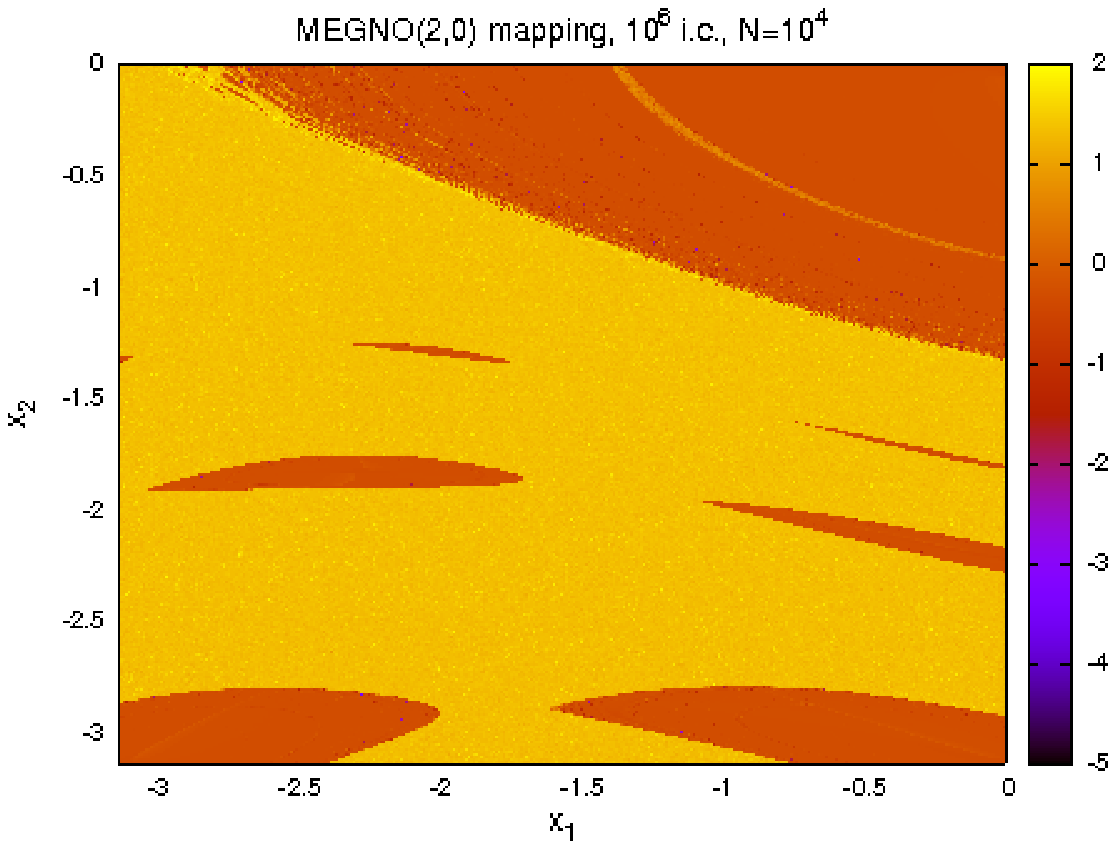}}&
\hspace{-5mm}\resizebox{63mm}{!}{\includegraphics{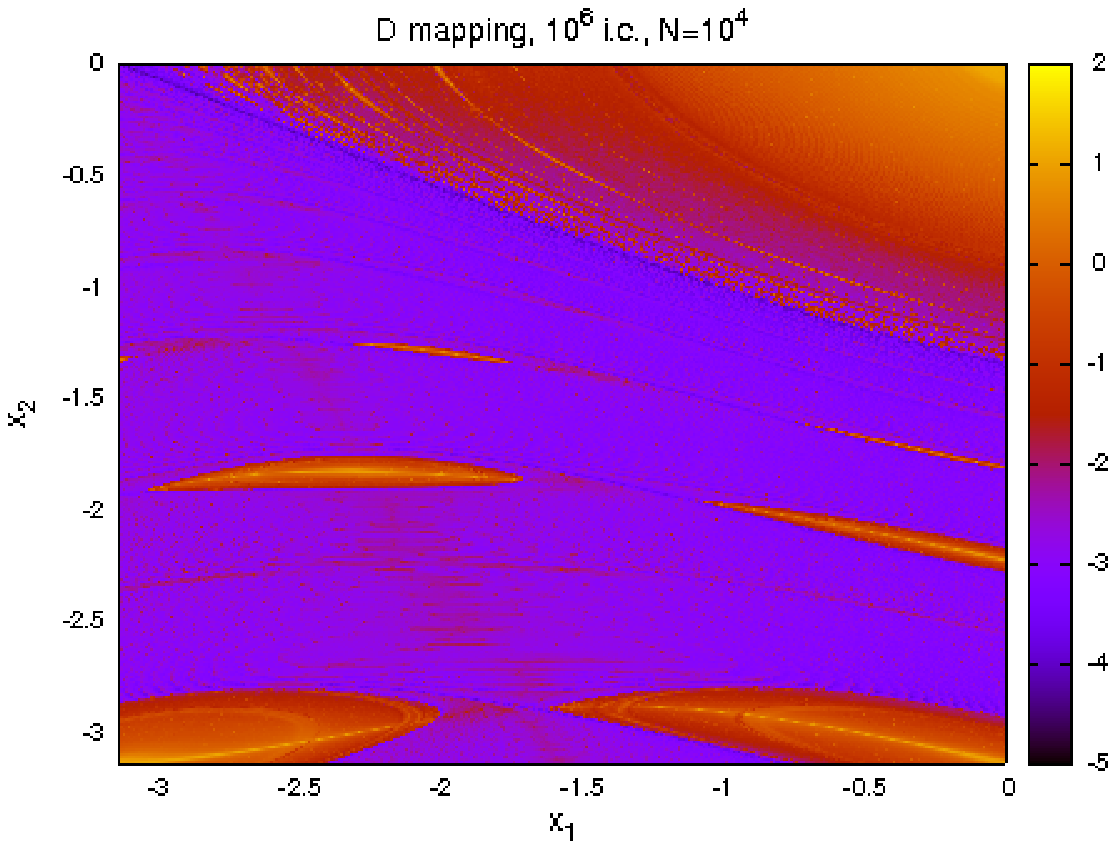}}
\end{tabular}
\caption{MEGNO(2,0) and \textit{D} mappings on color-scale plots composed of $10^6$ initial conditions, for $10^3$ (top panels) and $10^4$ (bottom panels) iterations. On the left, the MEGNO(2,0); on the right, the \textit{D}, in logarithmic scale.} 
\label{megnoandds1e6}
\end{center}
\end{figure}

On the left panels of Fig. \ref{megnoandds1e6}, we present the behavior of the MEGNO(2,0) for the vFSM, for $10^6$ initial conditions and after $10^3$ and $10^4$ iterations (on top and at the bottom, respectively). We observe the same characteristics shown by the LI or the RLI. The main difference is due to its theoretically fixed asymptotic threshold for the regular orbits. This particular characteristic of the MEGNO plays a key role when studying the time evolution of single orbits, as the regular motion and the stability levels can be easily identified by inspecting how the orbit converges to that fixed value (an advantage pointed out in many previous works; see Section \ref{Qualitative study-representative groups}). However, on working only with the final values, all the regular orbits have just one value, that of the threshold. Therefore, no further description is shown in any regular structure (e.g. in the secondary islands). The level of description in the case of the chaotic component is kept, since the MEGNO increases with $N$. 

The indicator \textit{D} (right panels of Fig. \ref{megnoandds1e6}) has a very noisy portrait for $10^3$ iterations (top panel) and it does not present a great improvement when multiplying $N$ by a factor of five or even ten (bottom panel). Only after $10^5$ iterations the \textit{D} clearly shows the stochastic layer inside the main stability island. Thus, the \textit{D} seems to delay a reliable description of the phase space portrait of the vFSM. 

\begin{figure}[ht!]
\begin{center}
\begin{tabular}{cc}
\hspace{-5mm}\resizebox{63mm}{!}{\includegraphics{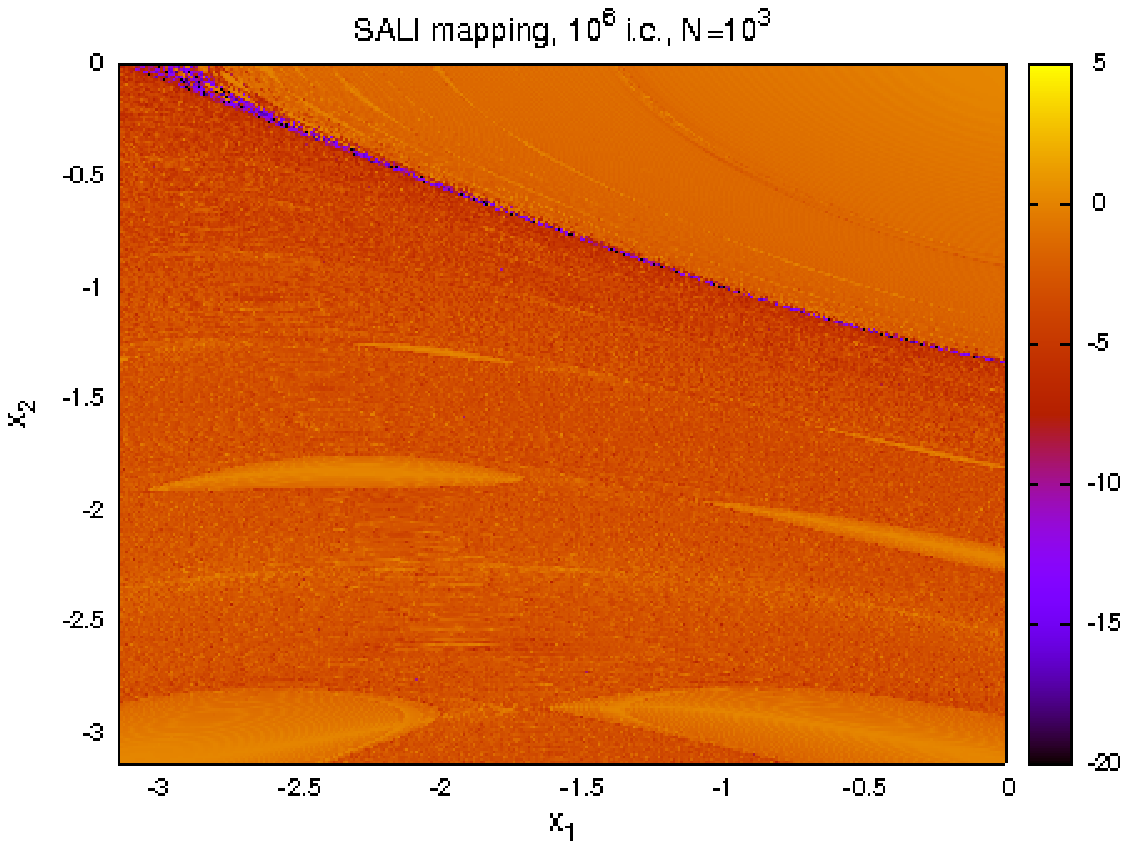}}&
\hspace{-5mm}\resizebox{63mm}{!}{\includegraphics{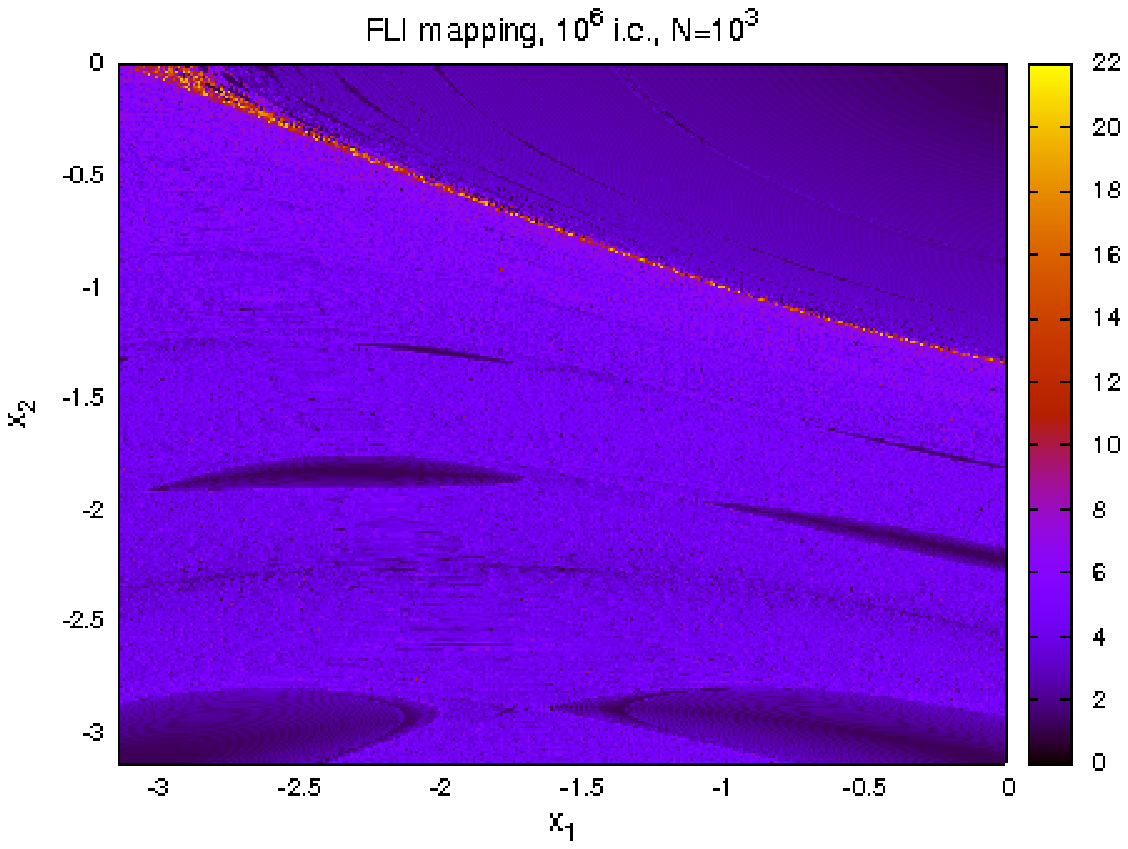}}\\ 
\hspace{-5mm}\resizebox{63mm}{!}{\includegraphics{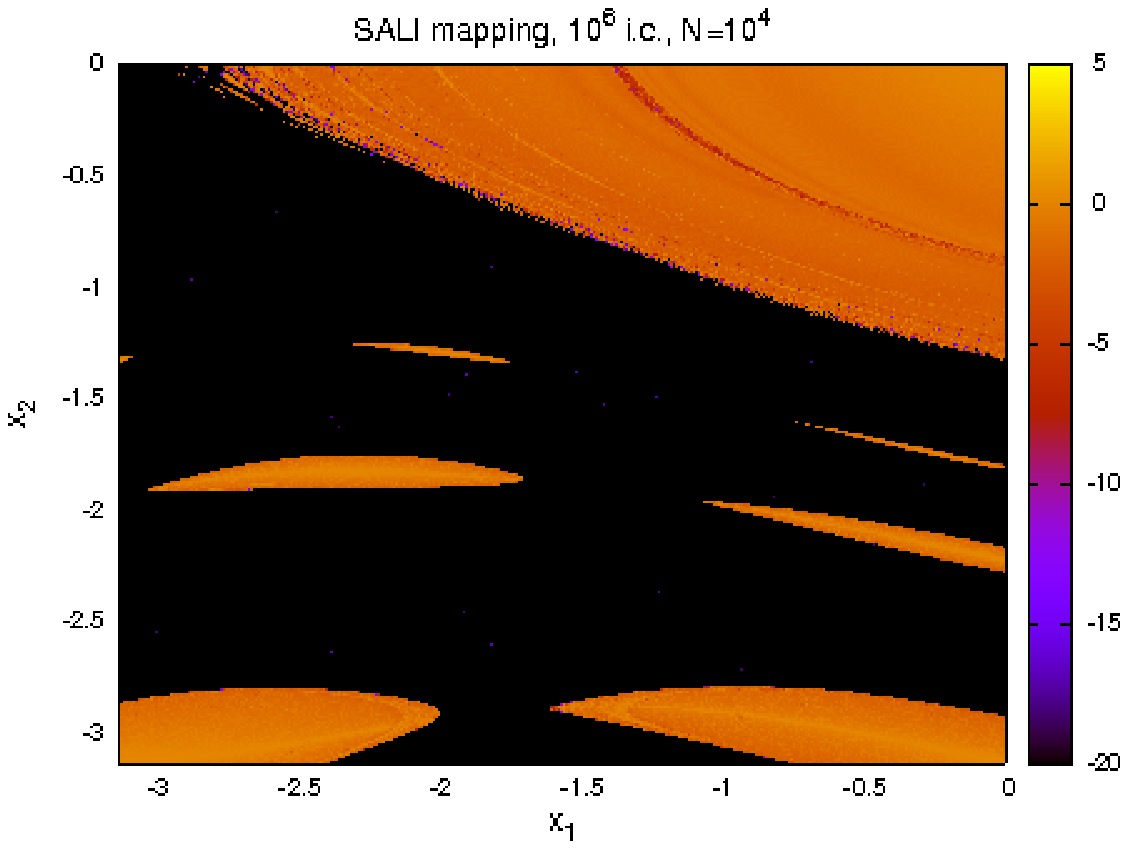}}&
\hspace{-5mm}\resizebox{63mm}{!}{\includegraphics{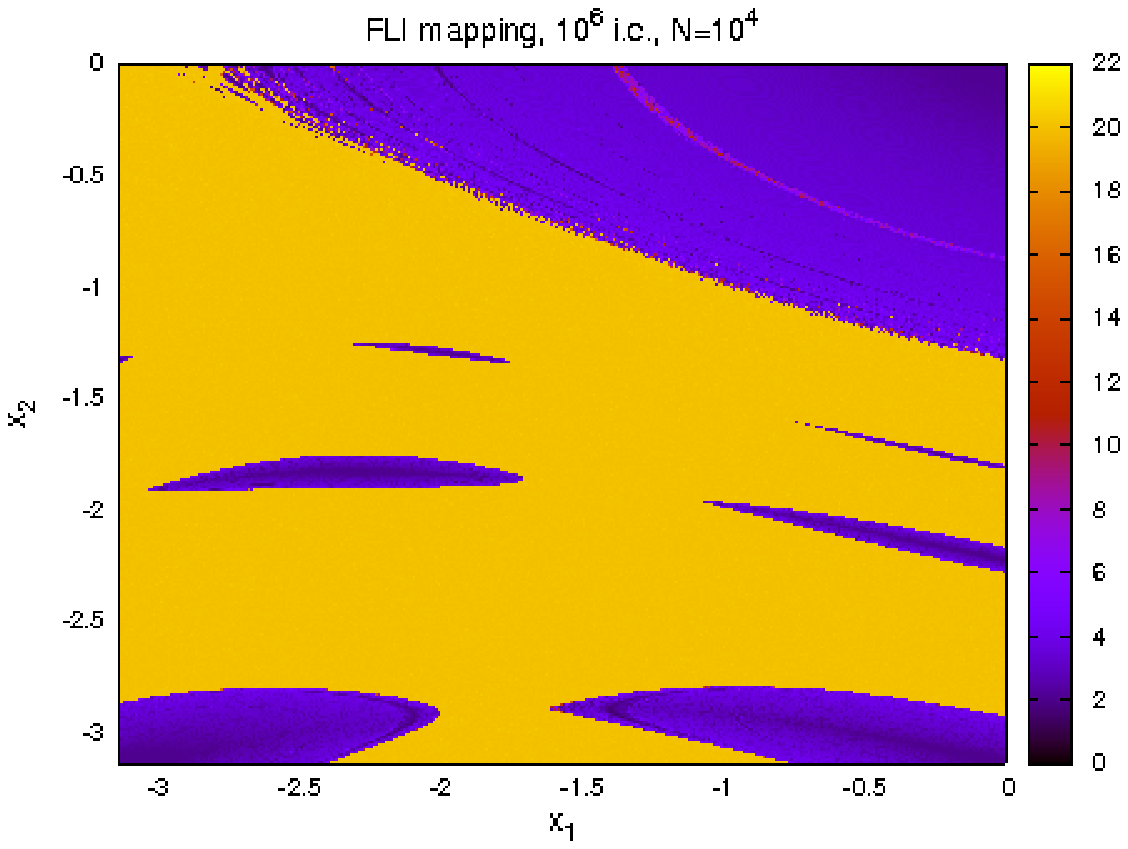}}
\end{tabular}
\caption{SALI and FLI mappings on color-scale plots of $10^6$ initial conditions, for $10^3$ (top panels) and $10^4$ (bottom panels) iterations. On the left, the SALI; on the right, the FLI, in logarithmic scale.} 
\label{saliandfli1e6}
\end{center}
\end{figure}

The SALI (left panels of Fig. \ref{saliandfli1e6}) and the FLI (right panels of Fig. \ref{saliandfli1e6}) show the same level of description as the other CI's. Although $10^3$ iterations do not suffice to have a clean phase space portrait (top panels of Fig. \ref{saliandfli1e6}), $5\times 10^{3}$ iterations do. However, as a result of the SALI and the FLI extremely high speed of convergence for the chaotic component, the logical consequence is the lack of information on the chaoticity levels. That is, most of their final values coincide with the saturation value when $N$ is large enough. Thus, the more suitable the selection of the $N$, the richer the description of the chaotic domain. By $10^4$ iterations, there is not a significant improvement (bottom panels of Fig. \ref{saliandfli1e6}) and with an $N$ ten times larger than the former, there is no difference at all. The SALI and the FLI reveal similar structures within the secondary islands.

\begin{figure}[ht!]
\begin{center}
\begin{tabular}{cc}
\hspace{-5mm}\resizebox{63mm}{!}{\includegraphics{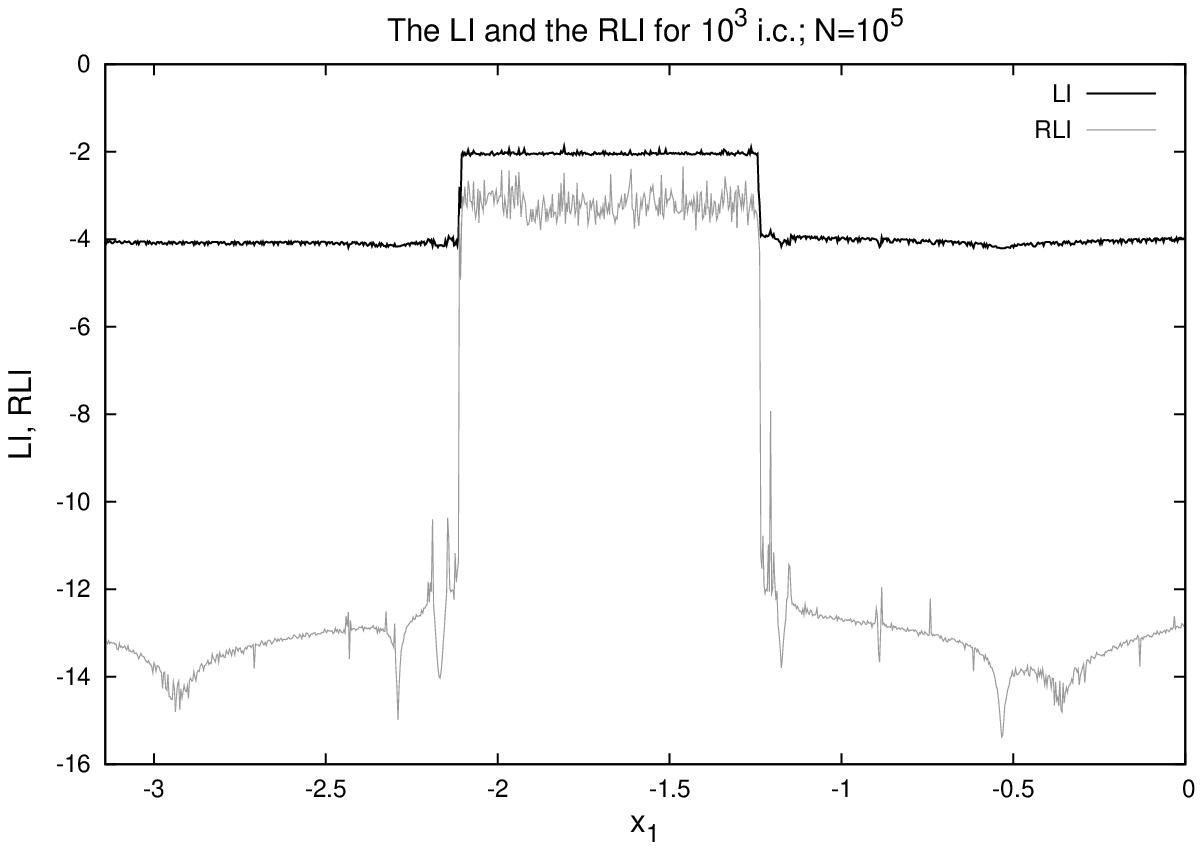}}&
\hspace{-5mm}\resizebox{63mm}{!}{\includegraphics{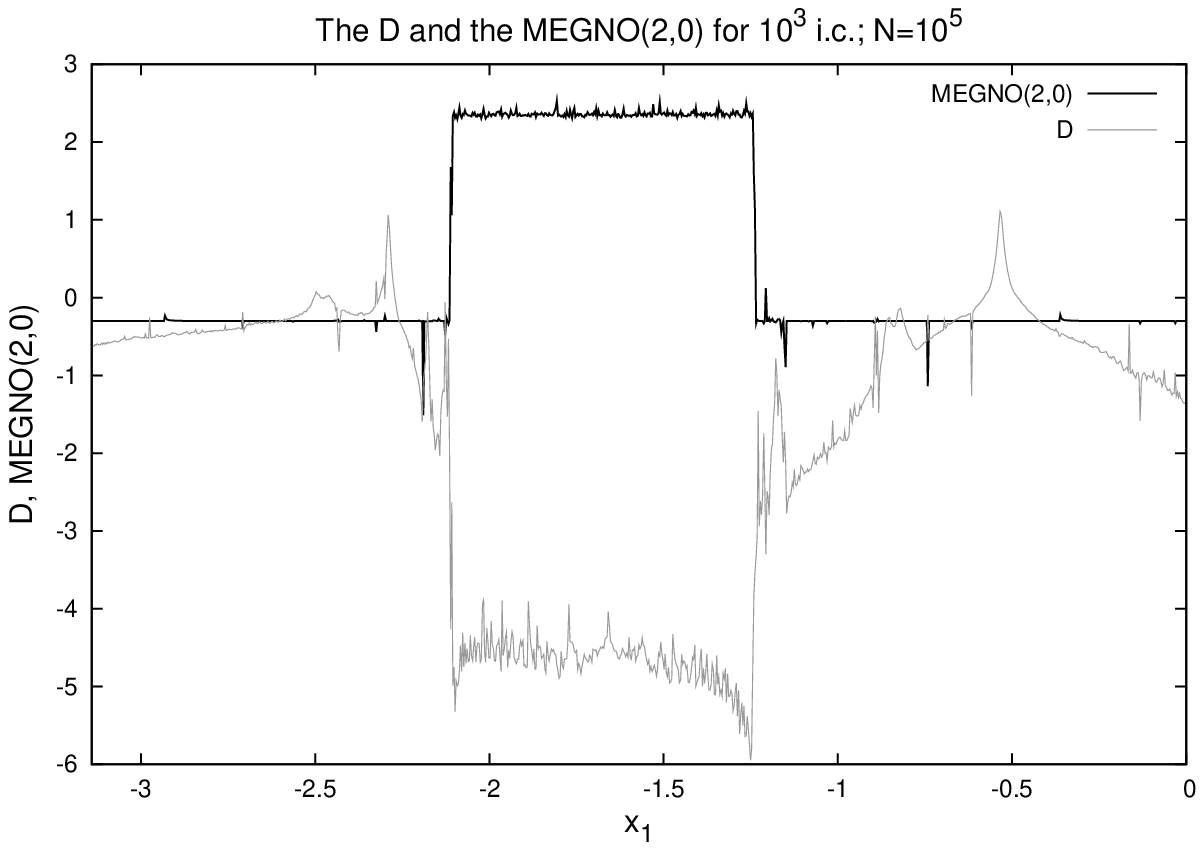}}\\ 
\hspace{-5mm}\resizebox{63mm}{!}{\includegraphics{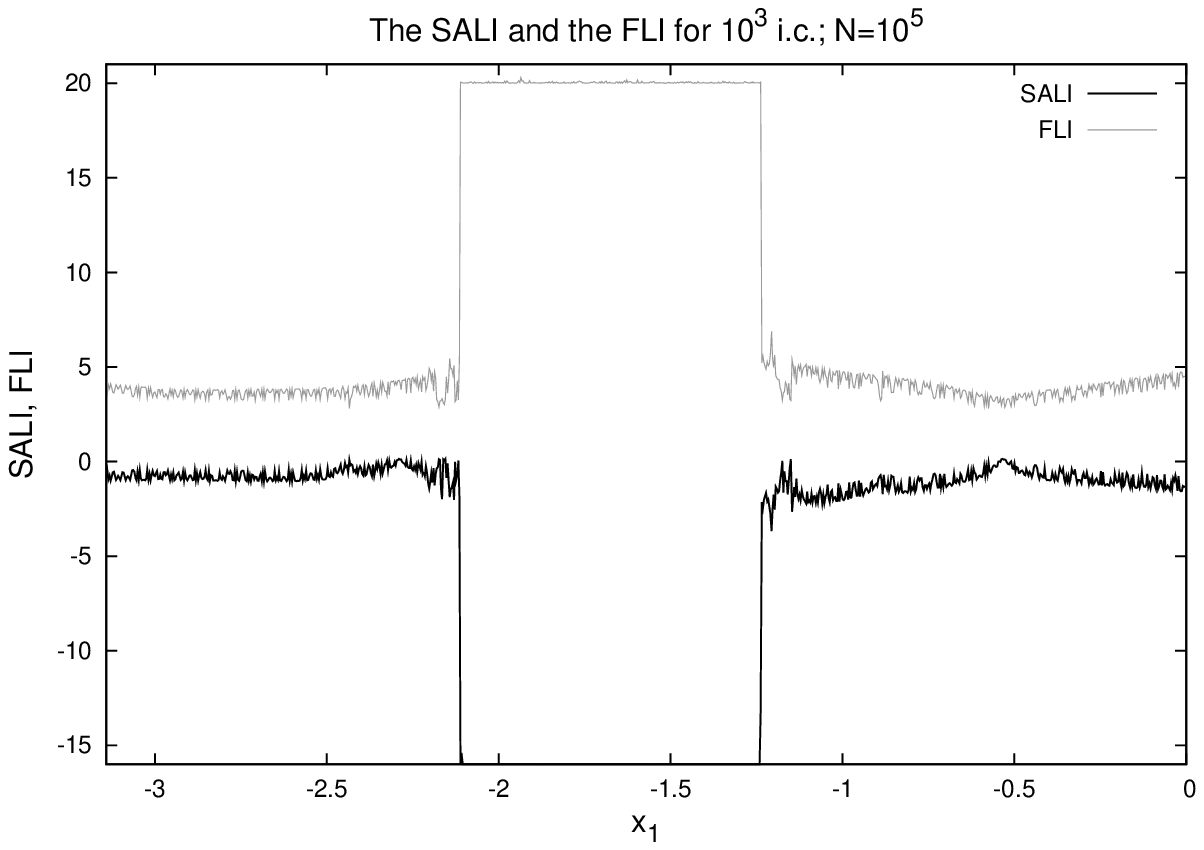}}&\\

\end{tabular}
\caption{The LI and the RLI (top left), the \textit{D} and the MEGNO(2,0) (top right) and the SALI and the FLI (bottom left) for $10^3$ equidistant initial conditions and for $10^5$ iterations along the line $x_2=-3$, in logarithmic scale.} 
\label{insideresonance}
\end{center}
\end{figure}

In Fig. \ref{insideresonance}, $10^3$ equidistant initial conditions $(x_1,\, x_2=-3,\, x_3=0.5,\, x_4=0)$ and $10^5$ iterations have been considered to describe the performance of the CIs along a line that crosses the high-order resonances mentioned earlier in this section. This figure clearly shows that the RLI and the \textit{D} are the indicators that best reveal some kind of structure inside these high-order resonances. The LI and the MEGNO(2,0) do not clearly show the structures. The SALI and the FLI give some extra information, but not as detailed as that of the RLI or the \textit{D}. Nevertheless, as it can be seen from the right panels of Fig. \ref{megnoandds1e6}, the \textit{D} is a rather noisy technique. Thus, the structures shown in Fig. \ref{insideresonance} by such indicator might be partially spurious (when comparing the structures shown by the RLI and the \textit{D} on the top left and right panels of Fig. \ref{insideresonance}, many important differences are observed). Therefore, the RLI proves to be the most accurate indicator to describe this large array of initial conditions. 
 
In conclusion, the time evolution is not efficient to analyze a large number of orbits and the appropriate way to gather information is through the final values of the CIs. In this scenario, among the CIs tested and within the vFSM, the RLI appears to be the most reliable indicator. It has a very sensitive resolving power and shows good performance in speed of convergence (partly due to a well-behaved threshold; see Section \ref{Thresholds study}).  

\section{Testing the CIs' main features under complex scenarios}\label{Extreme conditions}
To test the resolving power under complex conditions using the final values of every CI, we selected two regions of the vFSM that seem to be appropriate for the task. The first one is the small stochastic layer inside the main stability island (see the orange stripe on the bottom right panel of Fig. \ref{liandrli1e6}). The second one is the sticky region adjacent to such island. 

\subsection{The chaotic region inside the main stability island}\label{The chaotic region}
\begin{figure}[ht!]
\begin{center}
\begin{tabular}{cc}
\hspace{-5mm}\resizebox{63mm}{!}{\includegraphics{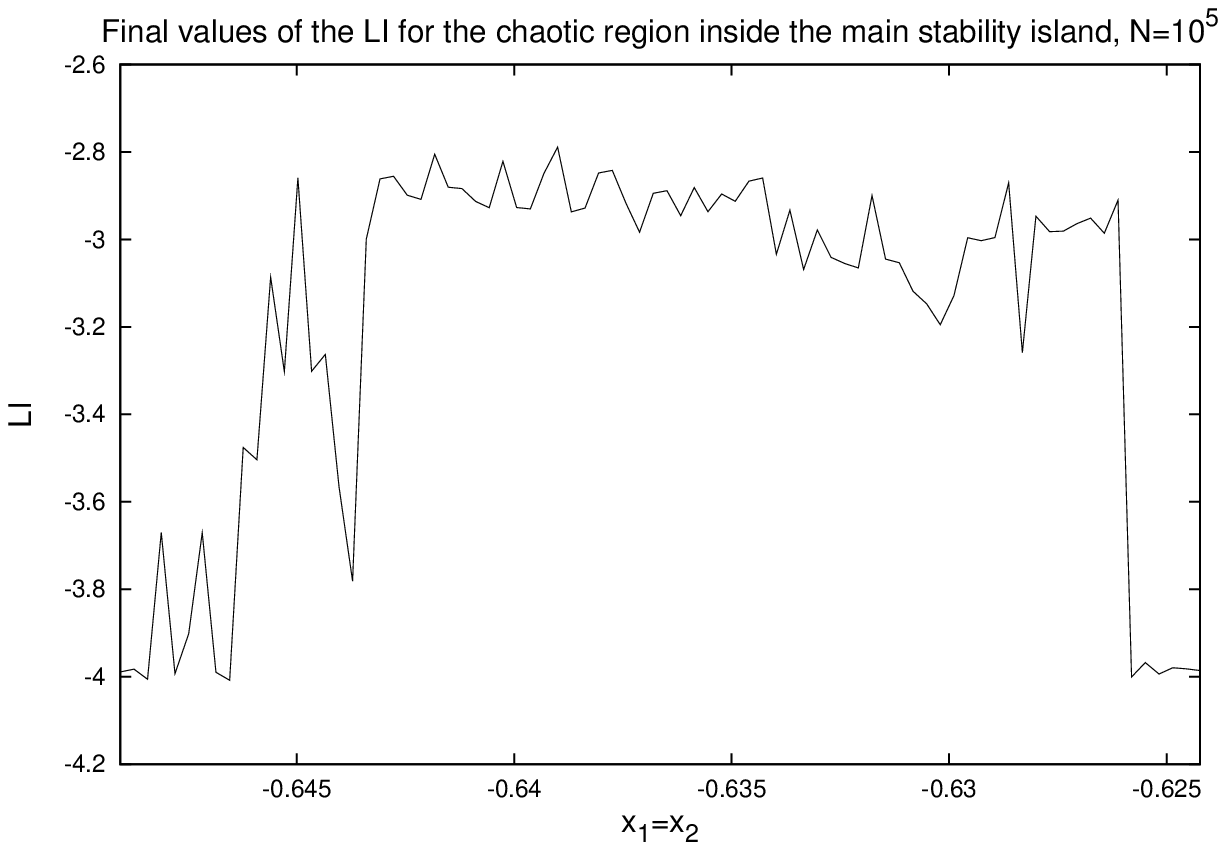}}& 
\hspace{-5mm}\resizebox{63mm}{!}{\includegraphics{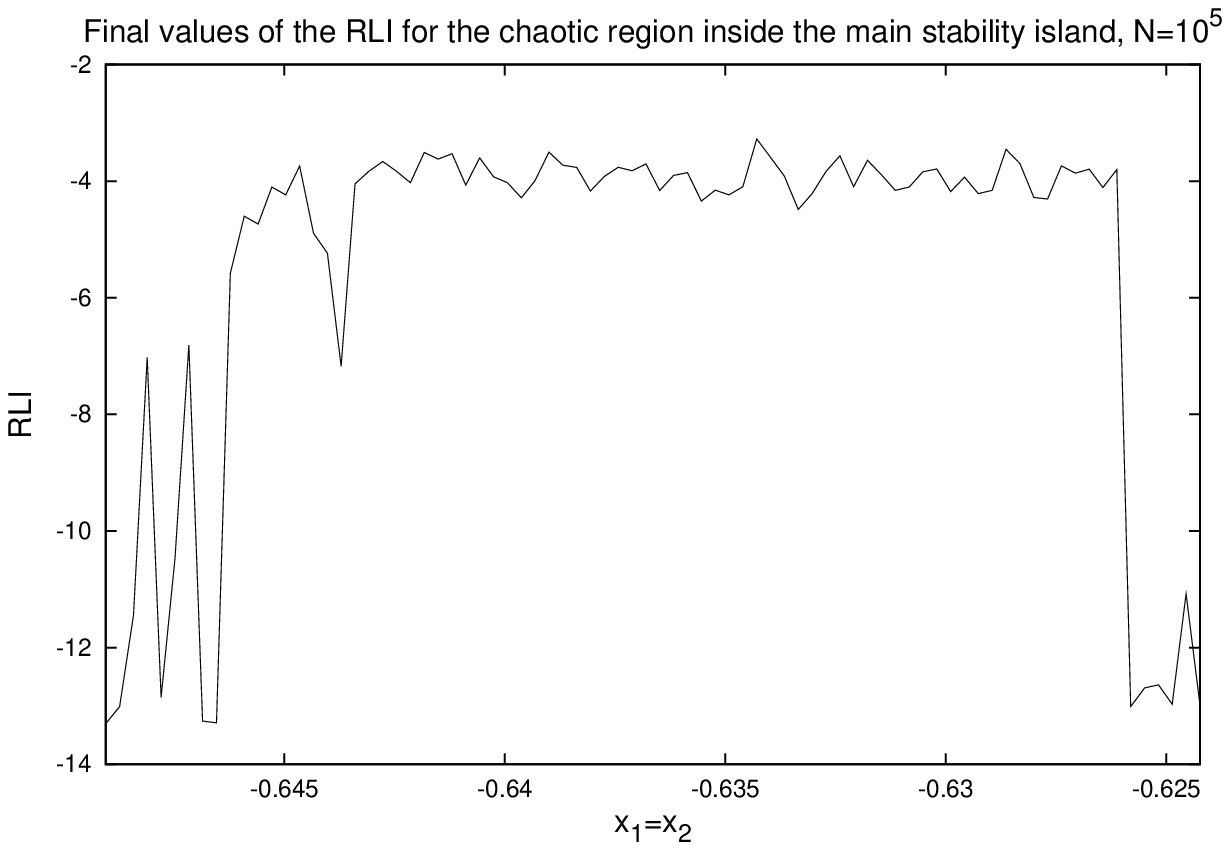}}\\
\hspace{-5mm}\resizebox{63mm}{!}{\includegraphics{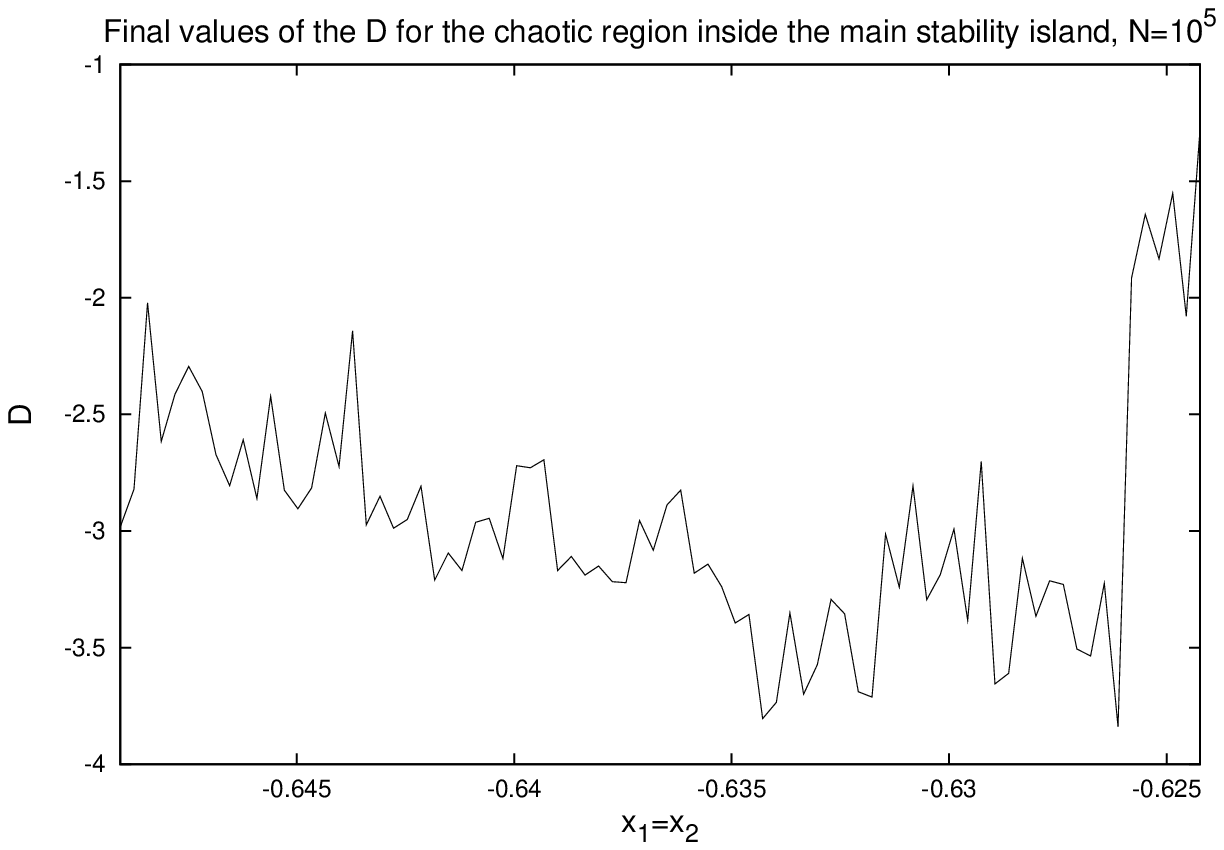}}& 
\hspace{-5mm}\resizebox{63mm}{!}{\includegraphics{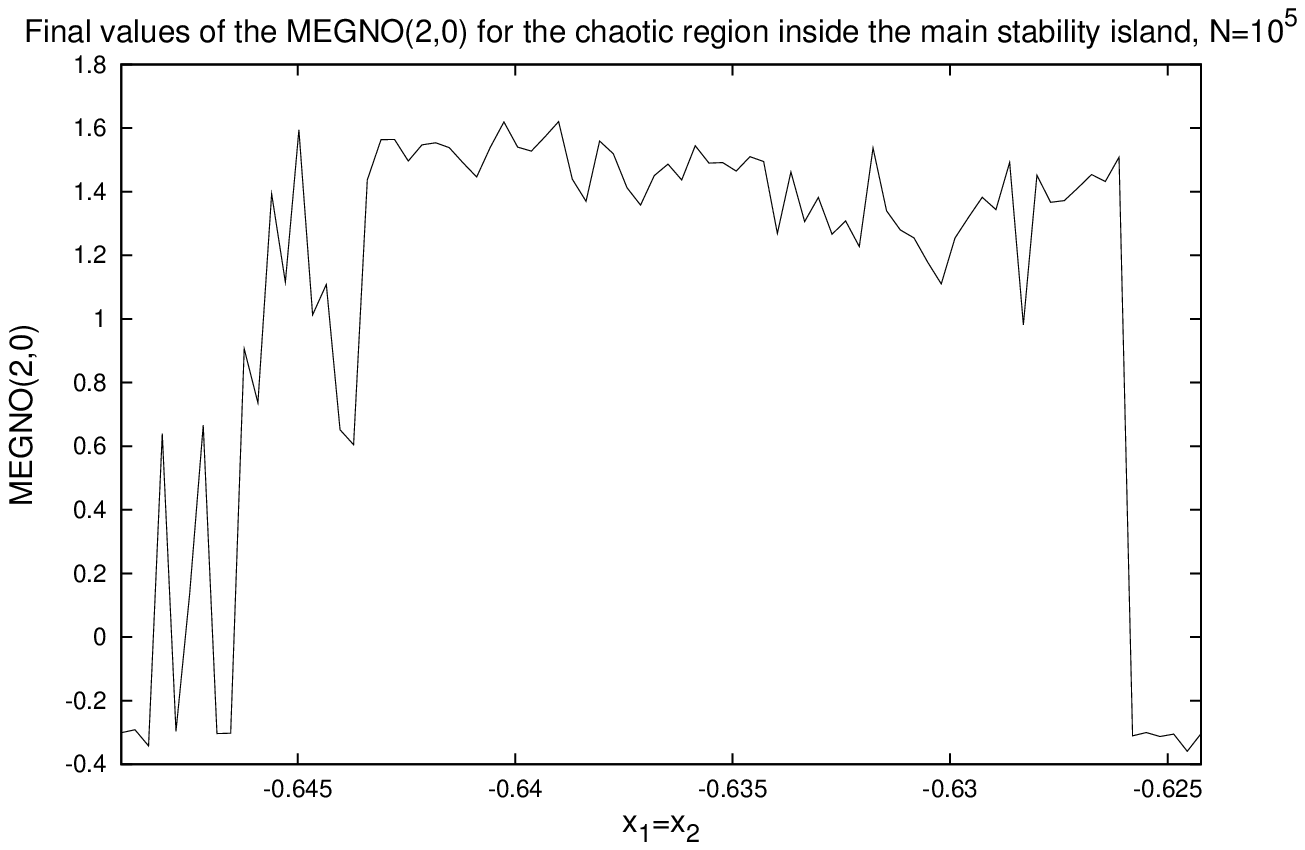}}\\
\hspace{-5mm}\resizebox{63mm}{!}{\includegraphics{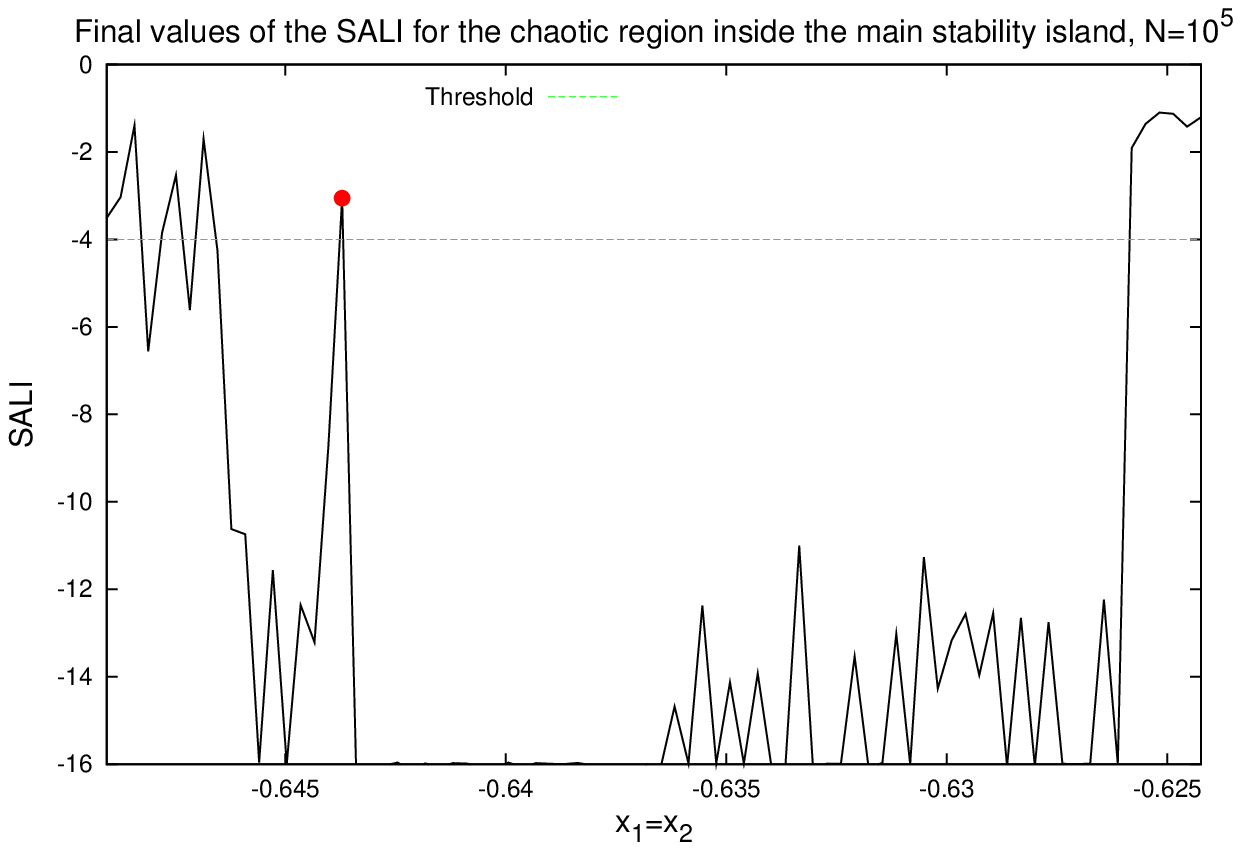}}& 
\hspace{-5mm}\resizebox{63mm}{!}{\includegraphics{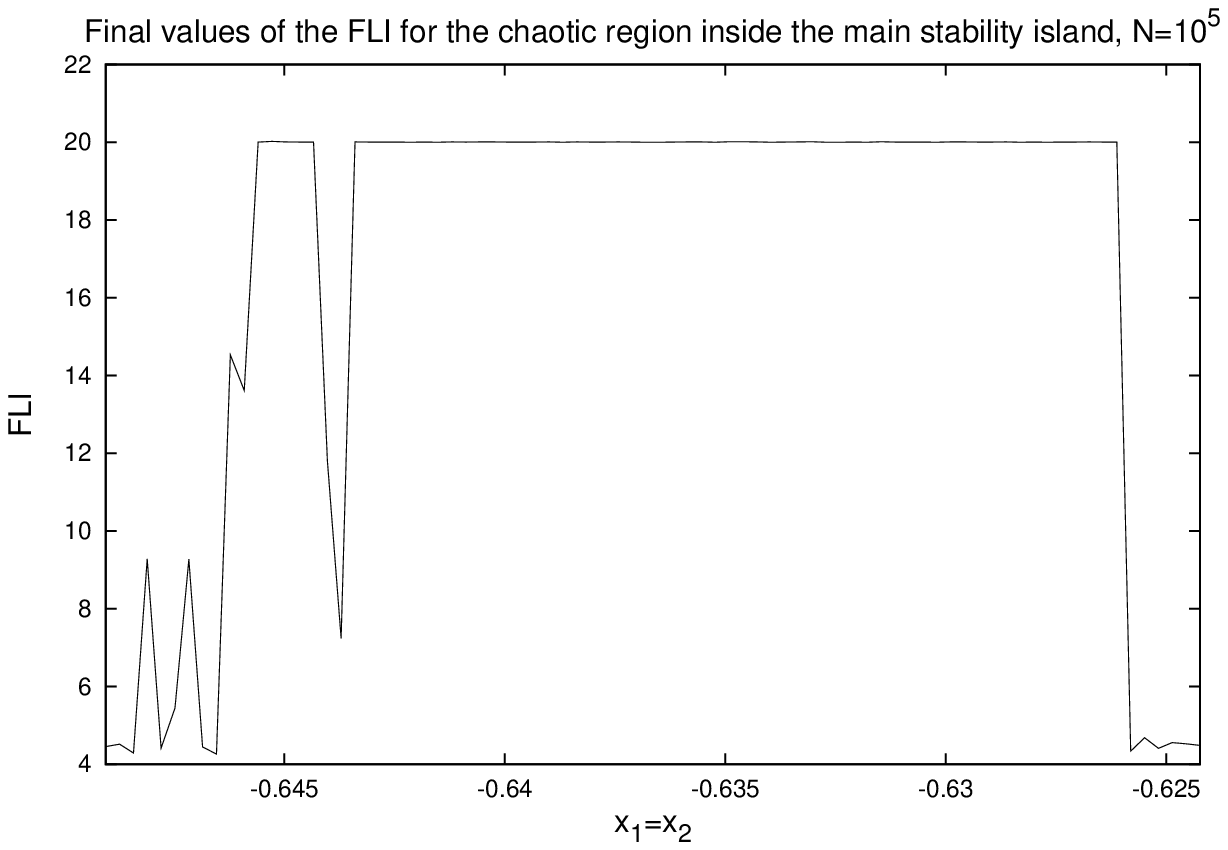}}
\end{tabular}
\caption{We show the profiles of the embedded chaotic region crossed along the identity line ($x_1=x_2$ and $x_1\in[-0.6490531,-0.6242345]$) for the different CIs. From top left to bottom right panels: the LI, the RLI, the \textit{D}, the MEGNO(2,0), the SALI and the FLI, respectively. In the case of the SALI (bottom left panel), the initial condition $x_1=x_2=-0.6437124$ (depicted in red color) and the threshold (depicted in green color) are included as well. In logarithmic scale.}
\label{chaosamp}
\end{center}
\end{figure}

In this section we will study the resolving power of the different CIs in the acknowledged chaotic region and its immediate surroundings. We take 80 orbits with initial conditions $x_1=x_2$ and $x_1\in[-0.6490531,-0.6242345]$ that cross the stochastic layer inside the main stability island along the identity line to have a first approach of the distribution of the motion. In Fig. \ref{chaosamp} we present such chaotic region and its neighborhood described by the CIs. The $N$ used for the experiment is $10^5$.

The LI, the RLI and the MEGNO(2,0) present a high level of coincidence in the description of the embedded chaotic zone (top left, top right and middle right panels of Fig. \ref{chaosamp}, respectively). However, the \textit{D} (middle left panel of Fig. \ref{chaosamp}) shows a structure that does not resemble the one shown by the other indicators. 

The FLI has no information about hyperbolicity levels (bottom right panel of Fig. \ref{chaosamp}) because of its high speed of convergence and the concomitant saturation value. The SALI, which has a similar rate of convergence, does display structures for the region. However, some of such structures might be spurious (bottom left panel of Fig. \ref{chaosamp}). The SALI has very small values for chaotic orbits, very close to the computer precision which might favor artificial formations. 

Regardless of the dependency of the SALI and the FLI on the saturation values (i.e. $10^{-16}$ for the SALI and $10^{20}$ for the FLI), it is possible to recover a measure of chaoticity by a quantity related to their final values. Let us consider the SALI, though the application can be extended to the FLI. A logical alternative to determine the hyperbolicity levels of chaotic orbits is the number of iterations with which the CI saturates, $N_{sat}$ (Skokos et al. 2007 have already used this idea to distinguish between chaotic and regular motions for the GALI$_3$ and the GALI$_4$). This is clearly seen when we follow the time evolution of the indicator for chaotic orbits. They reach the saturation value (the computer precision) at different numbers of iterations allowing us to distinguish one from the other. We make a plot using the two parameters that can be extracted from the SALI computation. The SALI does not saturate with the regular component and thus, the expected structures are successfully described by the SALI final values. Yet, in the case of the chaotic component, we retain the value $N_{sat}$ as a measure of the hyperbolicity levels since the SALI reaches the computer precision.
 
\begin{figure}[ht!]
\begin{center}
\begin{tabular}{cc} 
\hspace{-5mm}\resizebox{63mm}{!}{\includegraphics{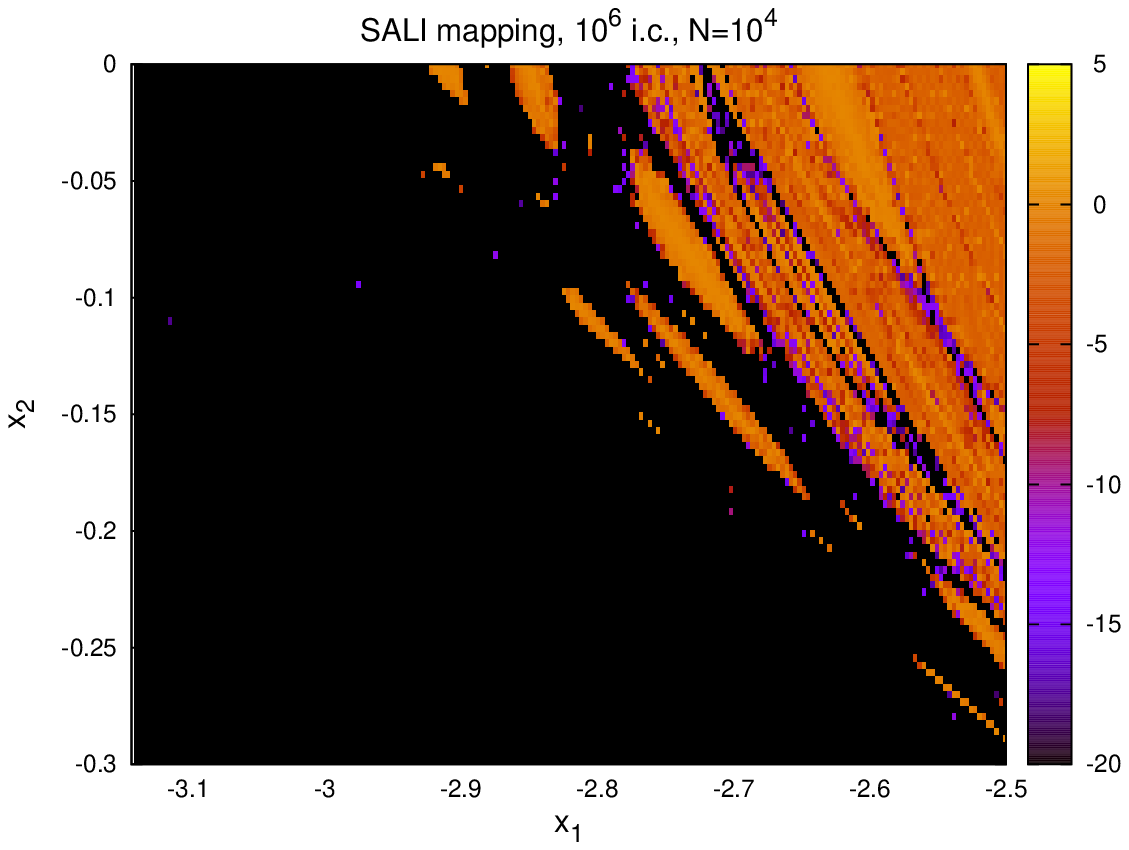}}& 
\hspace{-5mm}\resizebox{63mm}{!}{\includegraphics{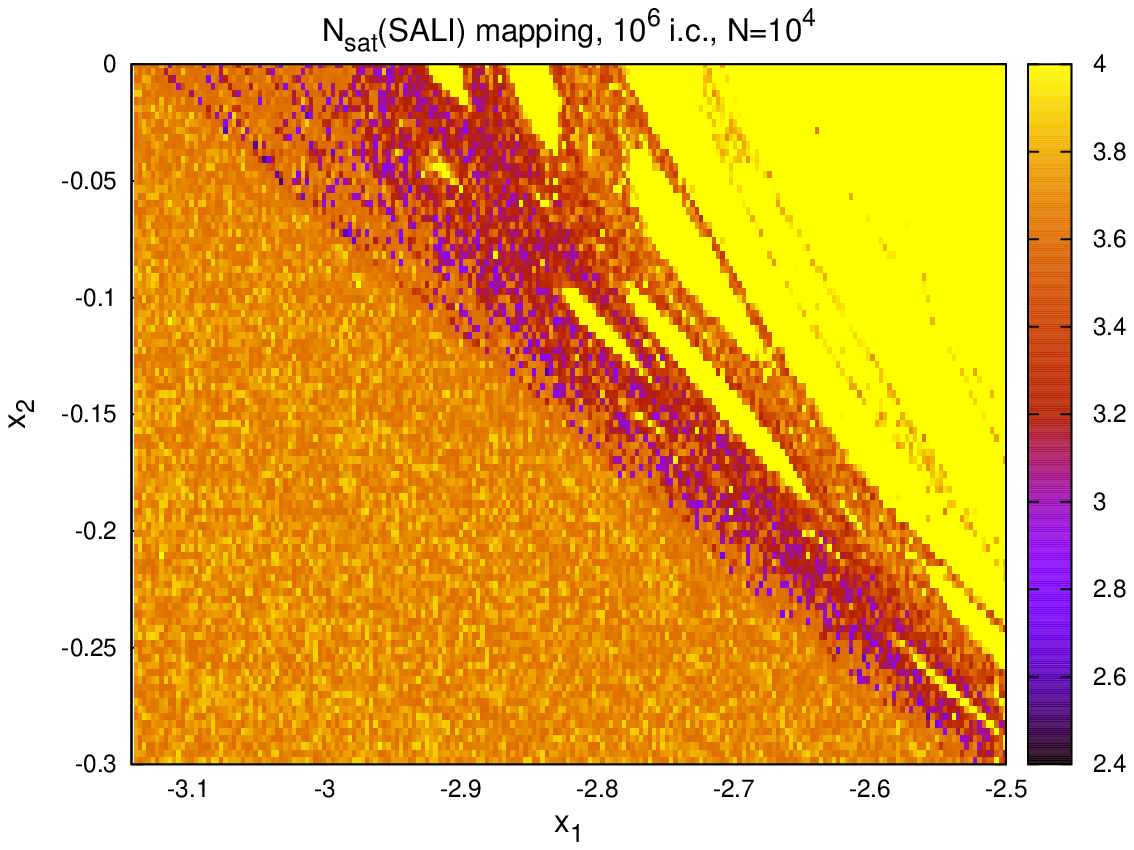}}
\end{tabular}
\caption{Left panel: SALI final values for $10^4$ iterations, zoom of the vFSM in the neighbourhodd of the main stability island (originally with $10^6$ i.c.), in logarithmic scale. Right panel: $N_{sat}$ for the same region, in logarithmic scale.}
\label{times10000}
\end{center}
\end{figure}

In Fig. \ref{times10000}, we present a zoom of the surroundings of the main stability island in the vFSM given by the final values of the SALI (left panel) and by the quantity $N_{sat}$ (right panel). Therefore, the regular component is described by the SALI final values and the chaotic component is described by the $N_{sat}$.

Considering both quantities obtained from a single computation of the SALI, namely, the final values and the $N_{sat}$, improve the analysis of statistical samples.
 
Finally, there is an orbit with initial condition $x_1=x_2=-0.6437124$ (the one marked on the bottom left panel in Fig. \ref{chaosamp}, located just above the threshold $10^{-4}$, which is also shown) that every technique but the SALI classifies as chaotic. This orbit has a high level of stickiness (which is seen by the time evolution curves), probably because of the proximity to a high-order resonance. Therefore, a rearrangement of the threshold of the SALI is needed. The time-independent threshold empirically given by Skokos et al. (2004) and used in this work, is an excellent estimation for statistical studies. However, such threshold should be carefully chosen when dealing with small samples in complex dynamics.

The results for the \textit{D} did not perform as well as those of the other indicators. 

\subsection{The sticky region adjacent to the main stability island}\label{The sticky region}
There is a region densely populated by sticky chaotic orbits adjacent to the main resonance. This kind of orbits are the most difficult to characterize by any indicator. In order to study the resolving power in this sticky region, we take a group of $\sim 760$ orbits that cross the surroundings of the main stability island along the identity line, within the interval (-1.03,-0.8) (see Table \ref{tableinicon3} for initial conditions of three representative orbits). The $N$ used for the analysis is $10^5$.

\begin{figure}[ht!]
\begin{center}
\begin{tabular}{cc}
\hspace{-5mm}\resizebox{63mm}{!}{\includegraphics{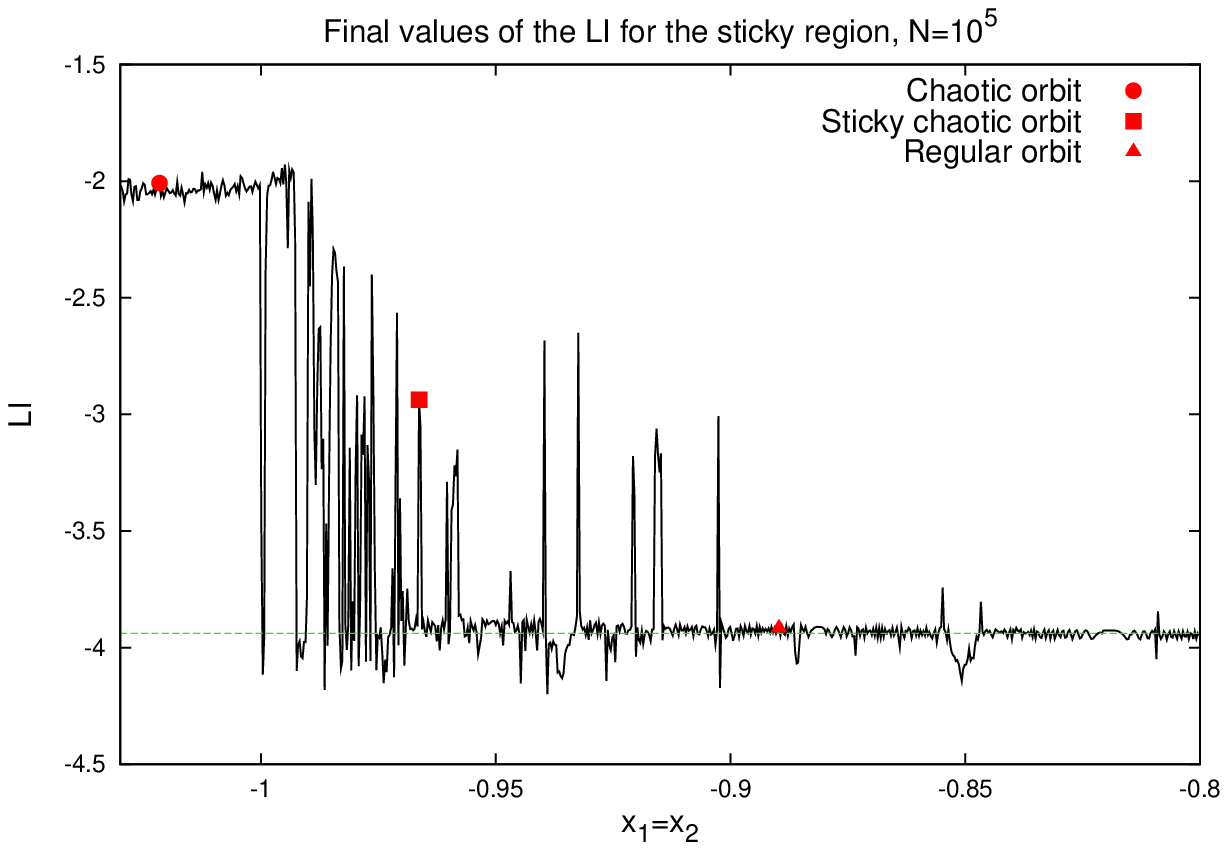}}& 
\hspace{-5mm}\resizebox{63mm}{!}{\includegraphics{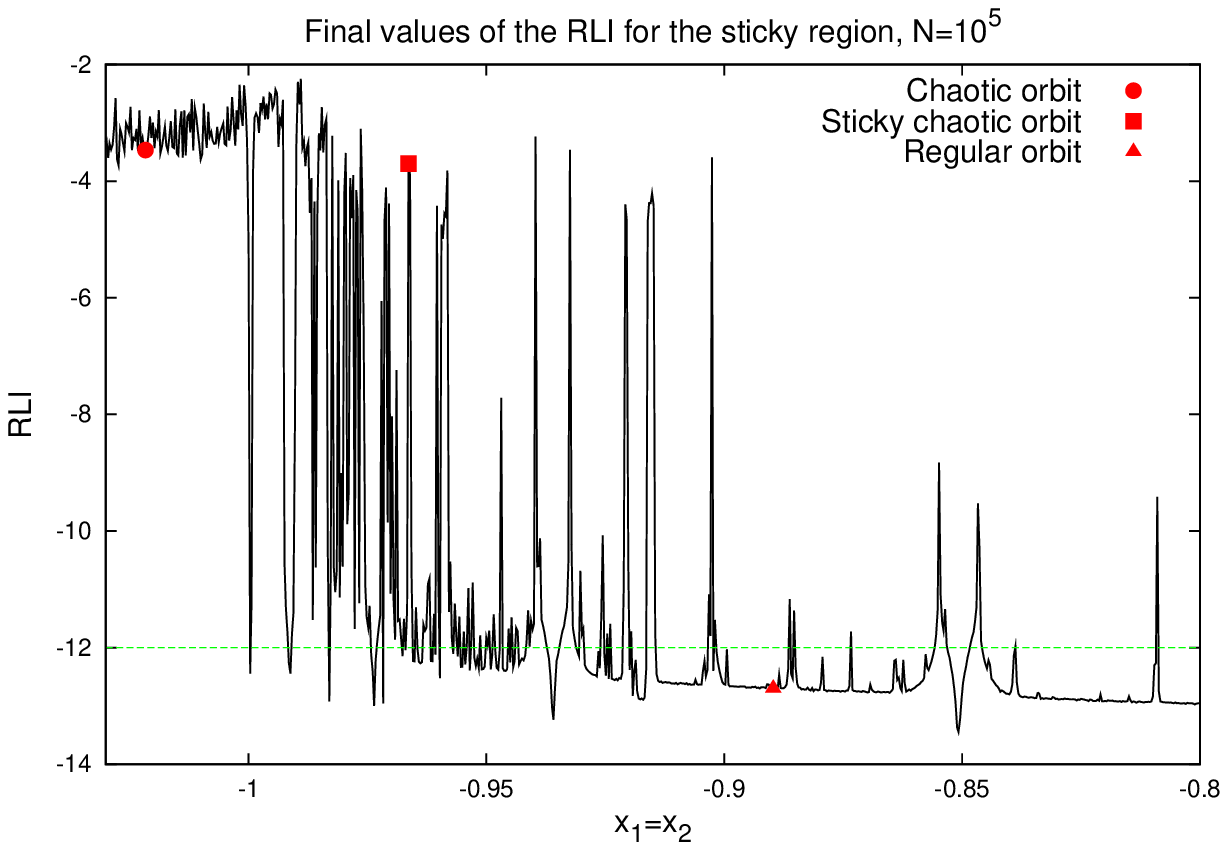}}\\
\hspace{-5mm}\resizebox{63mm}{!}{\includegraphics{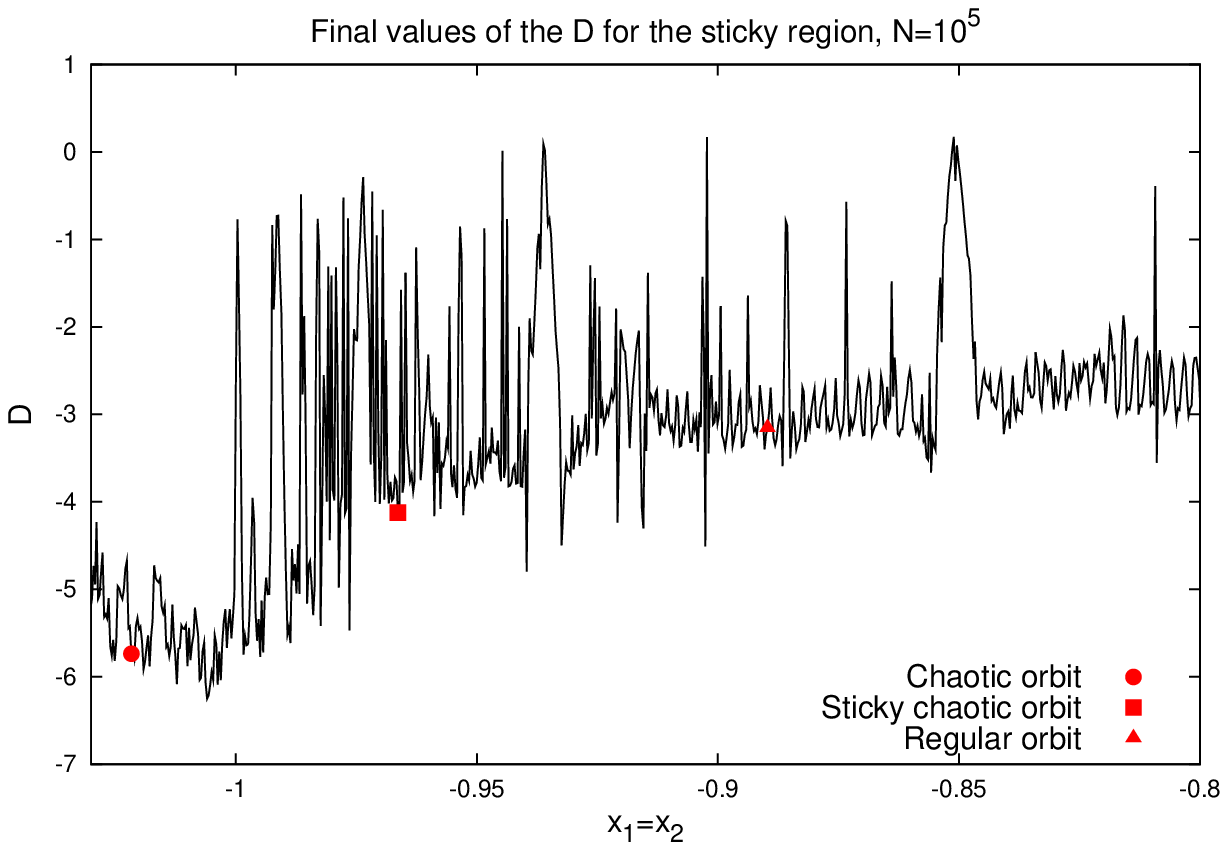}}& 
\hspace{-5mm}\resizebox{63mm}{!}{\includegraphics{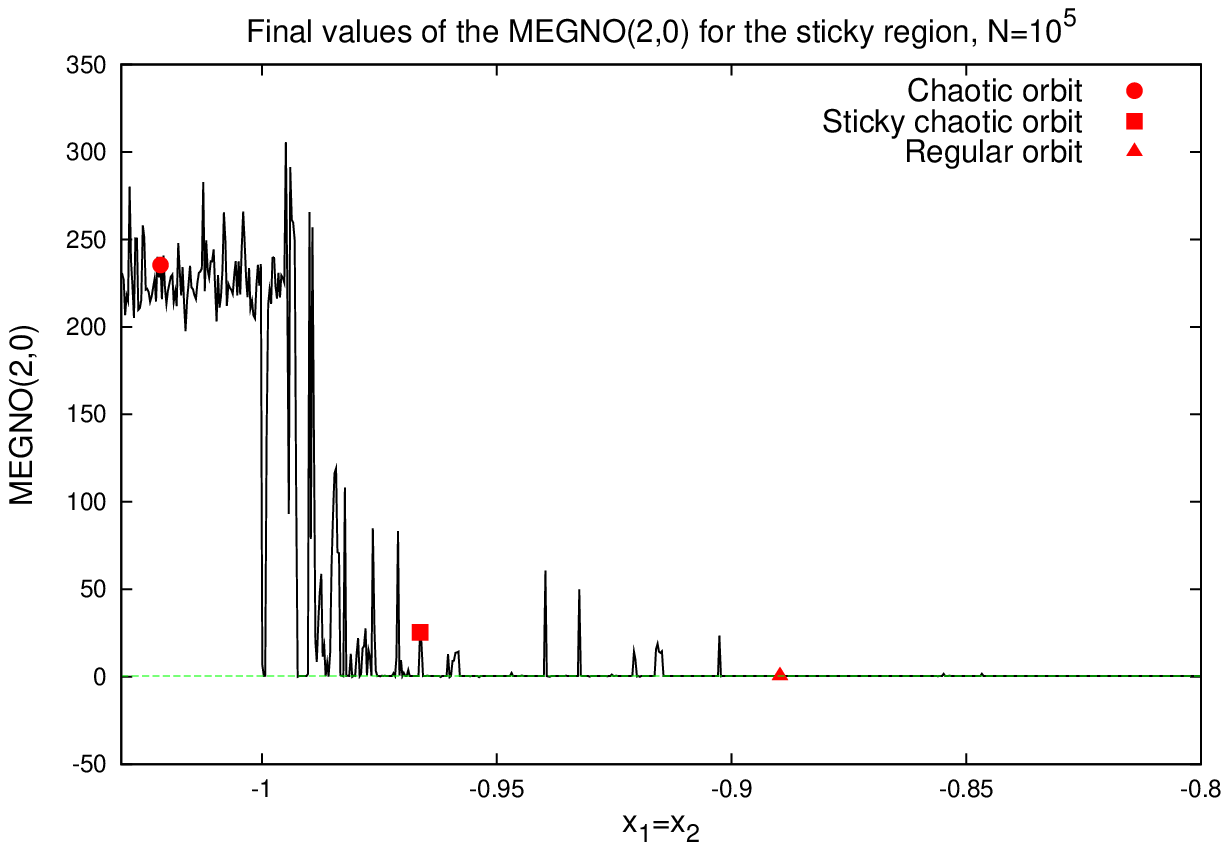}}\\
\hspace{-5mm}\resizebox{63mm}{!}{\includegraphics{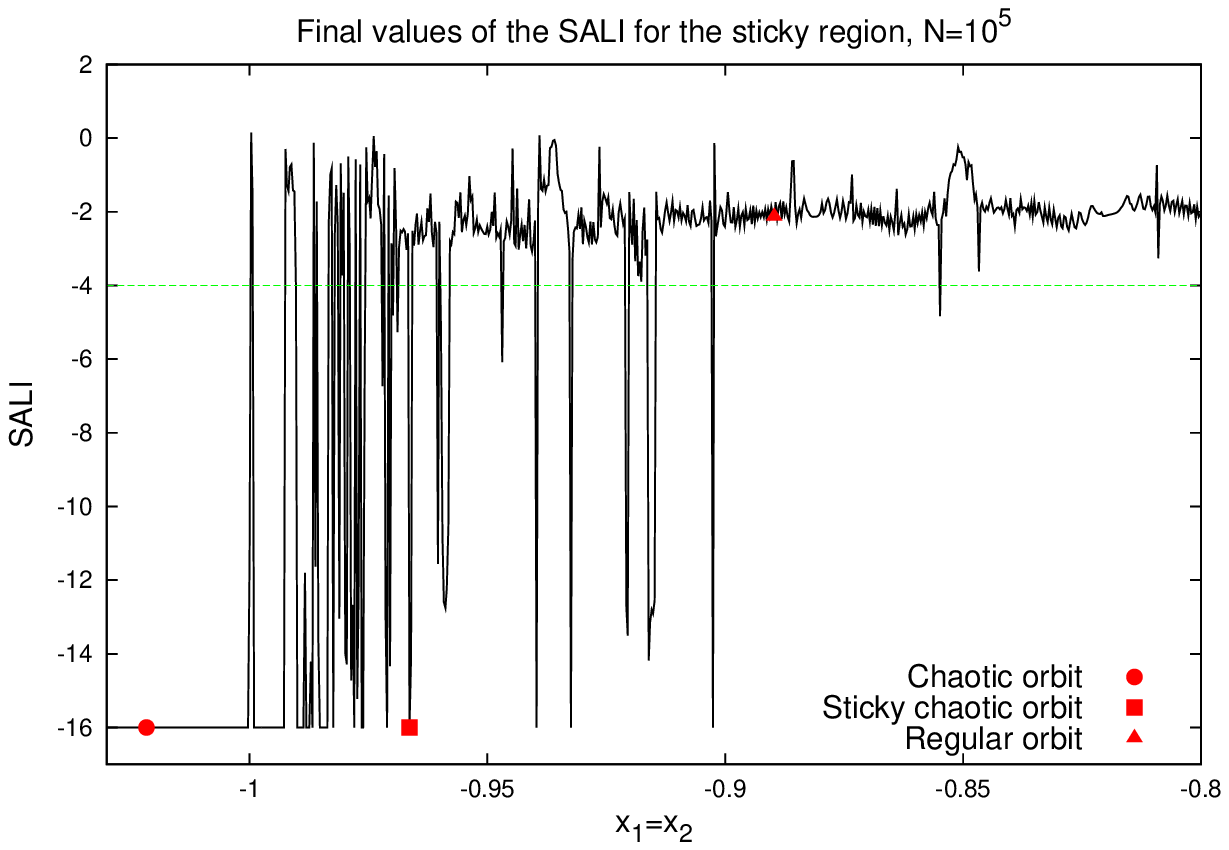}}& 
\hspace{-5mm}\resizebox{63mm}{!}{\includegraphics{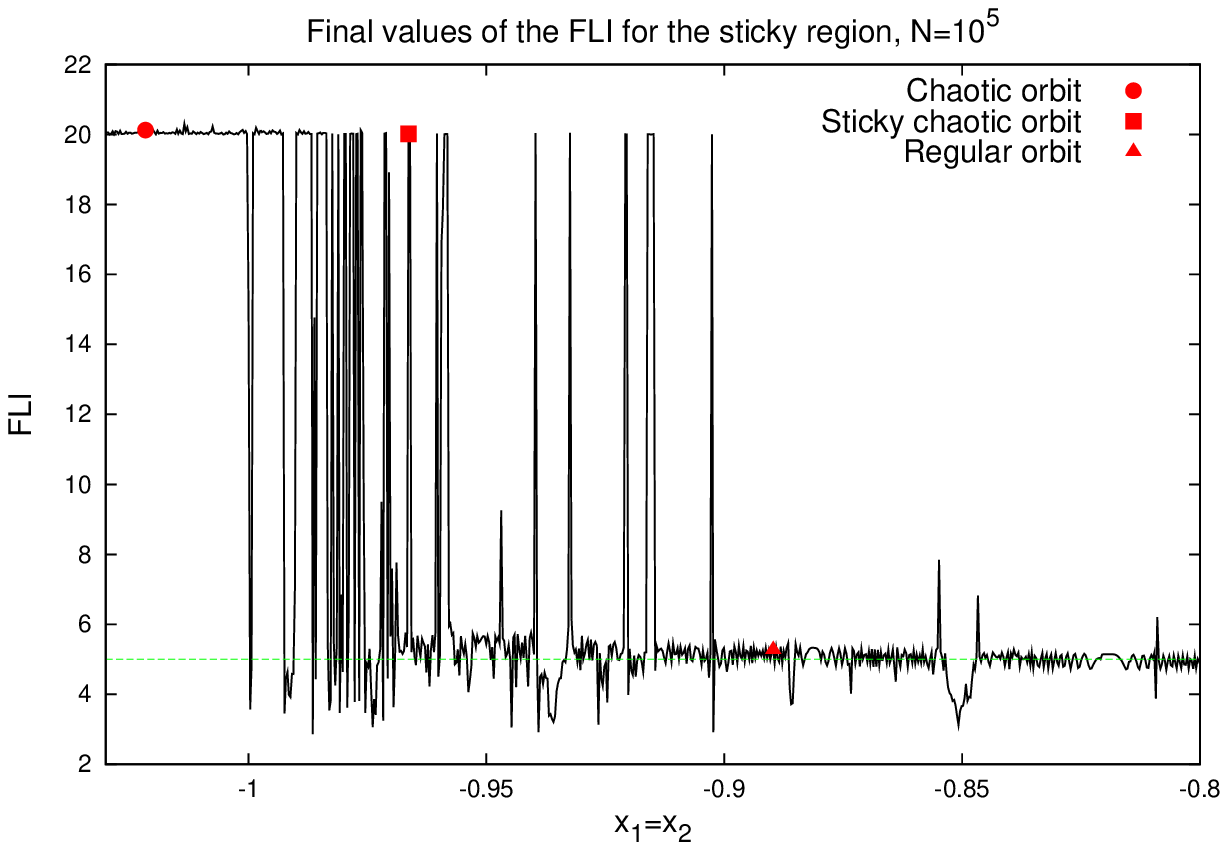}}
\end{tabular}
\caption{Zoom of the sticky region inside the interval (-1.03,-0.8). The known thresholds are depicted in green color. In logarithmic scale except for the MEGNO(2,0). See text for details.}
\label{stickyamp}
\end{center}
\end{figure}

In Fig. \ref{stickyamp}, the sticky region enclosed in the interval (-1.03,-0.8) is shown through the final values of the different CIs by $10^5$ iterations. We also pointed out the final values corresponding to the three selected orbits of Table \ref{tableinicon3}, two chaotic orbits (one of them sticky) and a regular orbit. The known thresholds are also included in the plots.

\begin{table}[!ht]\centering
\begin{tabular}{cccc}
\hline\hline  \vspace*{-2ex} \\ 
\emph{} Nature of the orbit & $x_1=x_2$ & $x_3$ & $x_4$ \vspace*{1ex} \\ 
\hline 
 Chaotic & -1.021646 & 0.5 & 0 \vspace*{1ex} \\
\hline 
 Chaotic (sticky) & -0.966354 & 0.5 & 0 \vspace*{1ex} \\
\hline 
 Regular & -0.8896991 & 0.5 & 0 \vspace*{1ex} \\
\hline\hline \vspace*{-4ex} 
\end{tabular}
\vspace{3mm}
\caption{Table of initial conditions for a group of orbits in the region adjacent to the main resonance.}
\label{tableinicon3}
\end{table}

We observed that the peaks and valleys shown in Fig. \ref{stickyamp} correspond to stochastic layers and resonances (the phase space portraits confirm this fact). Then, all indicators have good performances while revealing, globally, the phase space structure of the region. 

The description of the LI is presented on the top left panel of Fig. \ref{stickyamp}. The final values of the orbits inside the chaotic sea are close to the value $10^{-2}$ (on the left side of the panel where we find the selected chaotic orbit). The sticky region is shown as a complex structure of alternate peaks and valleys (a difference between the chaotic and the sticky chaotic orbit can be clearly observed). And finally, on the right side of the panel, the quasi constant value around $10^{-4}$ (consistent with the associated treshold though an empirical adjustment would be advisable) corresponds to the regular orbits inside the main resonance (where we have the representative regular orbit). Some small resonances of high-order are also shown in this region (e.g. $x_1=x_2\sim -0.85$).

We present the description of the RLI on the top right panel of Fig. \ref{stickyamp}. The main difference with the LI is the range of the values of the CI. The values of the LI lie between $10^{-4}$ and $10^{-2}$. In the case of the RLI, this interval is extended from values lower than $10^{-12}$ (which is the preferred threshold for the RLI to separate chaotic from regular motion) to above $10^{-3}$ because of the indicator's high speed of convergence. This fact might seem appropriate to unzip the information within the interval, but the high speed of convergence also decreases the quality of information in the chaotic component. Some sticky and chaotic orbits have similar final values which hide the different levels of hyperbolicity (such is the case of the representative chaotic and sticky chaotic orbits), this is not the case for the LI. However, the RLI has a good performance and quickly reveals the global characteristics of the system.  

As for the \textit{D}, the classification for the highlighted orbits (middle left panel of Fig. \ref{stickyamp}) coincides with the results obtained with the other indicators, but the spectra of final values is not so clean as for the other CIs. The reason is that the \textit{D} has a lower ability to separate instability levels. The distinction of the sticky region is rather confuse and the regular component associated to the main resonance (right side of the panel) shows big oscillations from one orbit to another.

The MEGNO(2,0)'s results agree with those of the LI about the chaotic and the sticky chaotic orbits (middle right panel of Fig. \ref{stickyamp}). In the case of the regular component, the orbits have values very close to the fixed threshold of $0.5$, due to the  MEGNO asymptotic behavior. The levels of stability that can be revealed by the MEGNO time evolution (see Section \ref{Qualitative study-representative groups}) are completely erased using only the final values, so no sub-structures are seen inside the high-order resonances. 

The SALI and the FLI achieve similar profiles (bottom left panel of Fig. \ref{stickyamp} for the SALI and bottom right panel of Fig. \ref{stickyamp} for the FLI). The chaotic orbits in the chaotic sea, on the left side of the corresponding figures, reach the corresponding saturation values, i.e. the level of accuracy of the computer: $10^{-16}$ for the SALI, or the value $10^{20}$ for the FLI. The final values of the SALI as well as the final values of the FLI for the chaotic orbit and the sticky chaotic orbit are the same because both orbits reach the related saturation value. For a better separation of the regular component, an adjustment of the threshold of the FLI might be appropriate (to see this, compare the location of the representative regular orbit from the corresponding threshold for both CIs).

When dealing with the final values of the sticky chaotic orbits the high speed of convergence becomes a disadvantage in their classification, because it allows them to saturate like chaotic orbits do (bottom panels of Fig. \ref{stickyamp}). Nevertheless, the $N_{sat}$ allows the SALI and the FLI to avoid such disadvantage. Some previous knowledge of the region is required for a better description of the system dynamics, e.g. through time evolution curves of a small sample of orbits to estimate the largest saturation time for chaotic orbits. Therefore, larger saturation times than those of such chaotic orbits would imply an amount of growing stickiness in the orbit.  

The LI and the MEGNO(2,0) show a simple way to distinguish sticky chaotic orbits by means of their final values while the RLI, the SALI and the FLI do not. But the SALI and the FLI can recover the information with the help of the time of saturation.

\section{Dependency on the parameters: the \textit{D} and the RLI}\label{A case of mild chaos}
The only CIs from the package that depend strongly on ``user-choice'' parameters are the \textit{D} and the RLI. The \textit{D} needs the arrangement of histograms to be computed, and the RLI needs an initial separation between the basis orbit and its ``shadow'' (S\'andor et al. 2004). Therefore, these dependencies deserve greater analysis.

In the articles of Voglis et al. (1999) and Skokos (2001), there is a discrepancy in the performance of the \textit{D} that is worthy of some review. They analyze a dynamical system comprising two coupled standard mappings where there exists a well-known case of weak chaos (orbit A3 as the authors called it in Voglis et al. 1999).  

The equations for the mapping are:

$$x'_1=x_1+x'_2$$
$$x'_2=x_2+\frac{K}{2\pi}\sin[2\pi\cdot x_1]-\frac{\beta}{\pi}\sin[2\pi(x_3-x_1)]$$
$$x'_3=x_3+x'_4$$
$$x'_4=x_4+\frac{K}{2\pi}\sin[2\pi\cdot x_3]-\frac{\beta}{\pi}\sin[2\pi(x_1-x_3)]$$
with $K=3$, and $\beta=0.1$ or $\beta=0.3051$.

Herein, the orbits A2 (regular) and A3 (weakly chaotic, see Voglis et al. 1999 for details) are studied by means of the whole package of CIs. We use the same initial deviation vectors and the same $N$ used by Voglis et al. (1999). 

\begin{figure}[ht!]
\begin{center}
\begin{tabular}{cc}
\hspace{-5mm}\resizebox{63mm}{!}{\includegraphics{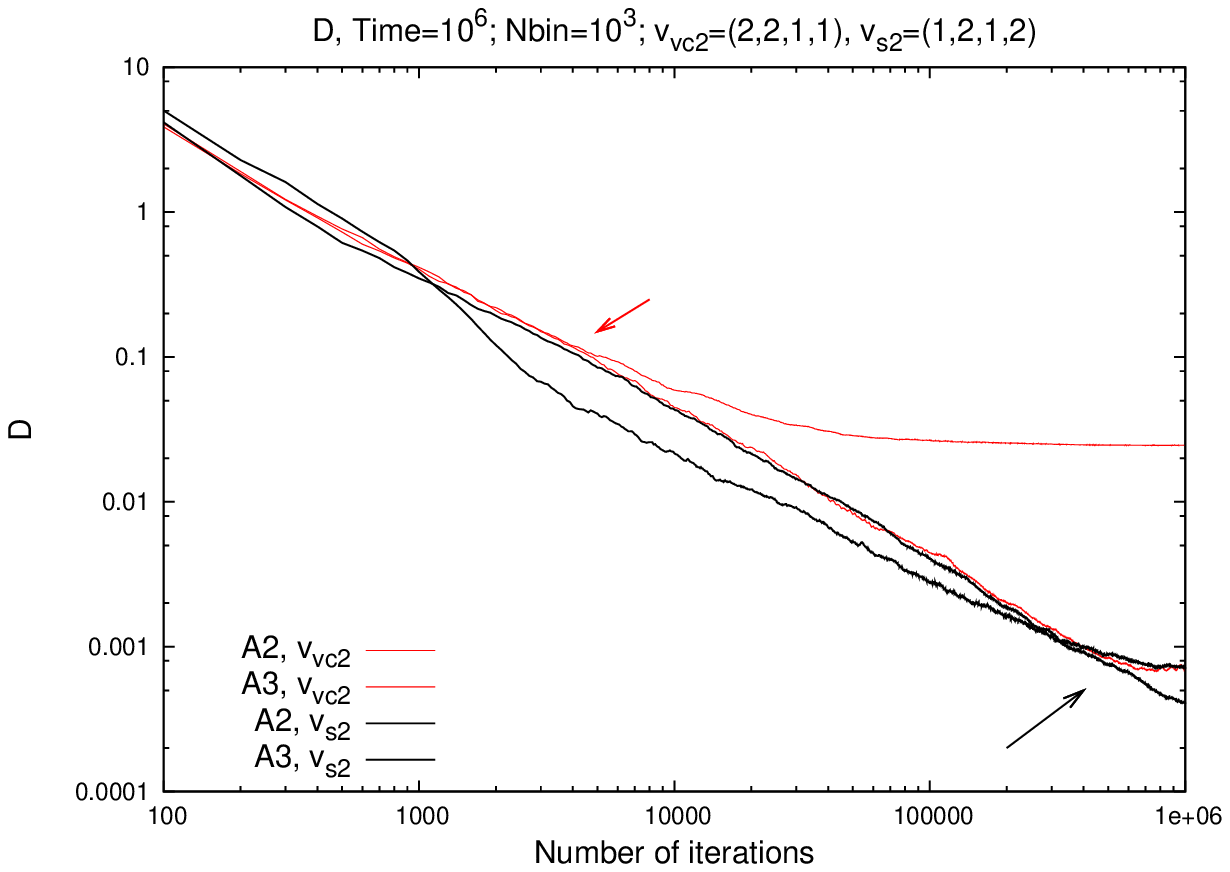}}&
\hspace{-5mm}\resizebox{63mm}{!}{\includegraphics{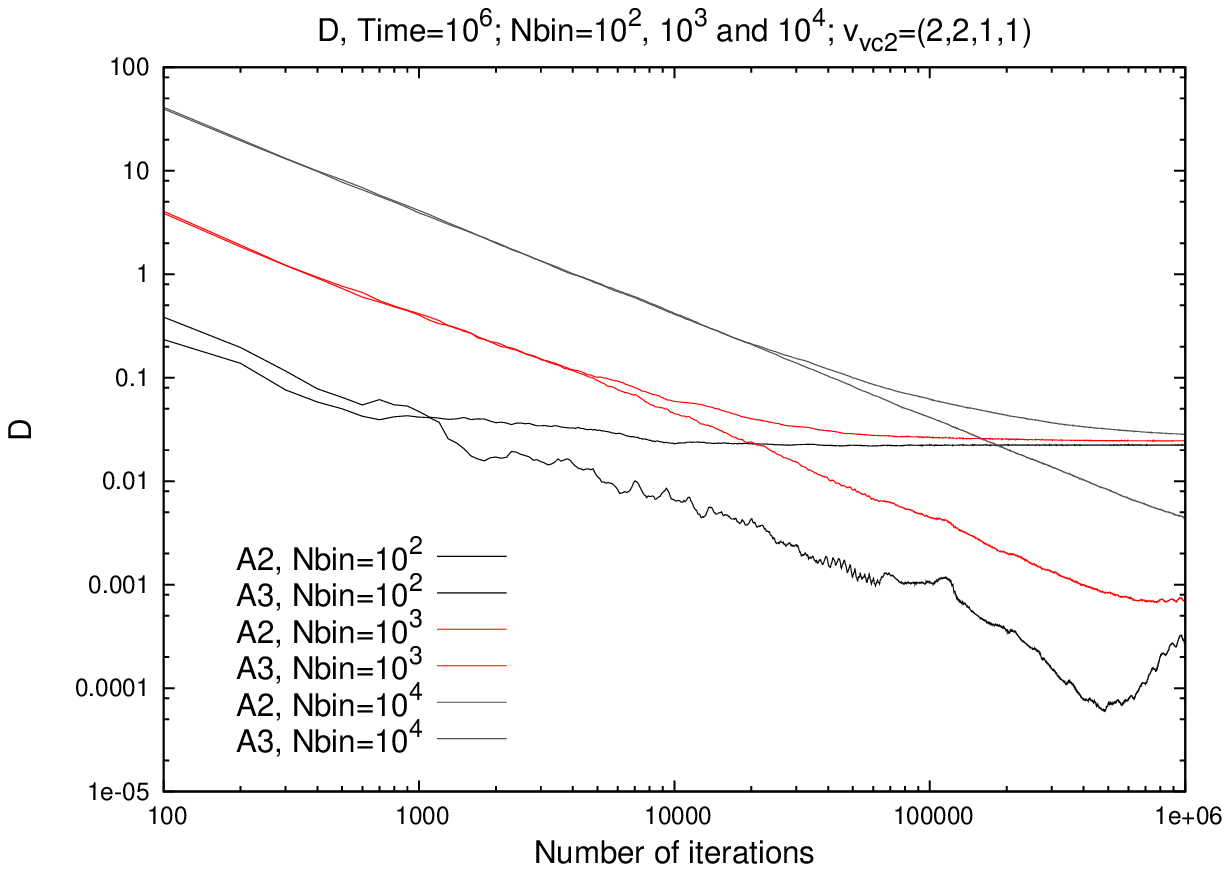}}
\end{tabular}
\caption{Behaviors of the \textit{D} using the pair of initial deviation vectors selected by Voglis et al. (1999) ($(1,1,1,1)$ and $v_{vc2}=(2,2,1,1)$) depicted in red color and Skokos (2001) ($(1,1,1,1)$ and $v_{s2}=(1,2,1,2)$) depicted in black color, for orbits A2 and A3. The separation of the orbits is depicted with an arrow (left panel). Variation of the \textit{D} with the $Nbin$ parameter for orbits A2 and A3 on the right panel.}
\label{problemas}
\end{center}
\end{figure}

The result of the \textit{D} matches that provided by the authors; i.e., it seems to be the fastest indicator to distinguish the weakly chaotic orbit from the regular orbit. However, in Skokos (2001), where the author studies the same set of orbits, the conclusion is rather different (compare Fig. 4 from Voglis et al. 1999 and bottom left panel of Fig. 8 from Skokos 2001). On the left panel of Fig. \ref{problemas}, we have the orbits A2 and A3 computed with two pairs of initial deviation vectors, namely, the pair taken by Voglis et al. (1999): (1,1,1,1) and (2,2,1,1); and the one chosen by Skokos (2001): (1,1,1,1) and (1,2,1,2). It is easy to see that the particular choice of Voglis et al. (1999) improves the distinction of the orbits by the \textit{D} in almost two orders of magnitude (notice the arrows that point out the separations corresponding to each choice of initial deviation vectors). However, for the purposes of our study, the best way to compare the performances of the indicators is to use a random choice of initial deviation vectors. In fact, on doing so, the \textit{D} has a similar rate of convergence to that of the rest of the CIs in the package.   

As we have seen, the \textit{D} seems to provide valuable information if some free parameters could be efficiently adjusted, e.g. the initial deviation vectors. Such adjustment is not easy and prior knowledge of the expected results is sometimes required. Therefore, it is important to quantify this sensitivity of the \textit{D} with some other parameters involved in its computation. 

The $Nbin$ is the number of cells used to build the histograms for the determination of the SSN and the \textit{D}. A high sensitivity to the variation of $Nbin$ is observed not only with the CPU times, but also on the right panel of Fig. \ref{problemas}. An earlier separation of the regular and weakly chaotic orbits A2 and A3, respectively, is reached when decreasing the $Nbin$ parameter. The smaller the $Nbin$ taken for the determination of the histograms, the faster the distinction between regular and chaotic motion is shown. We have more points in each cell; thus, the differences between the curves are amplified. The experiments carried out give us an idea that $10^2$ points per cell is a fair estimate for the $Nbin$ parameter (thus, e.g., for $10^5$ iterations, $Nbin=10^3$ is a reasonably value, and it is the one used to reproduce the results of Voglis et al. 1999 and Skokos 2001). 

The RLI has a free parameter also: the initial separation of the two orbits (see S\'andor et al. 2004). Therefore, it might be of interest to evaluate the sensitivity of the indicator to this parameter as we did with the \textit{D}. 

The initial separation does not significantly affect the RLI final values for chaotic orbits, but it does in the case of the regular component. 

The way the initial separation parameter affects the RLI final values for the regular orbits considered agrees with the analysis made by S\'andor et al. (2004), where they conclude that the correspondence is linear. However, other orbits may require a slightly different approach. 

\begin{figure}[ht!]
\begin{center}
\begin{tabular}{cc}
\hspace{-5mm}\resizebox{63mm}{!}{\includegraphics{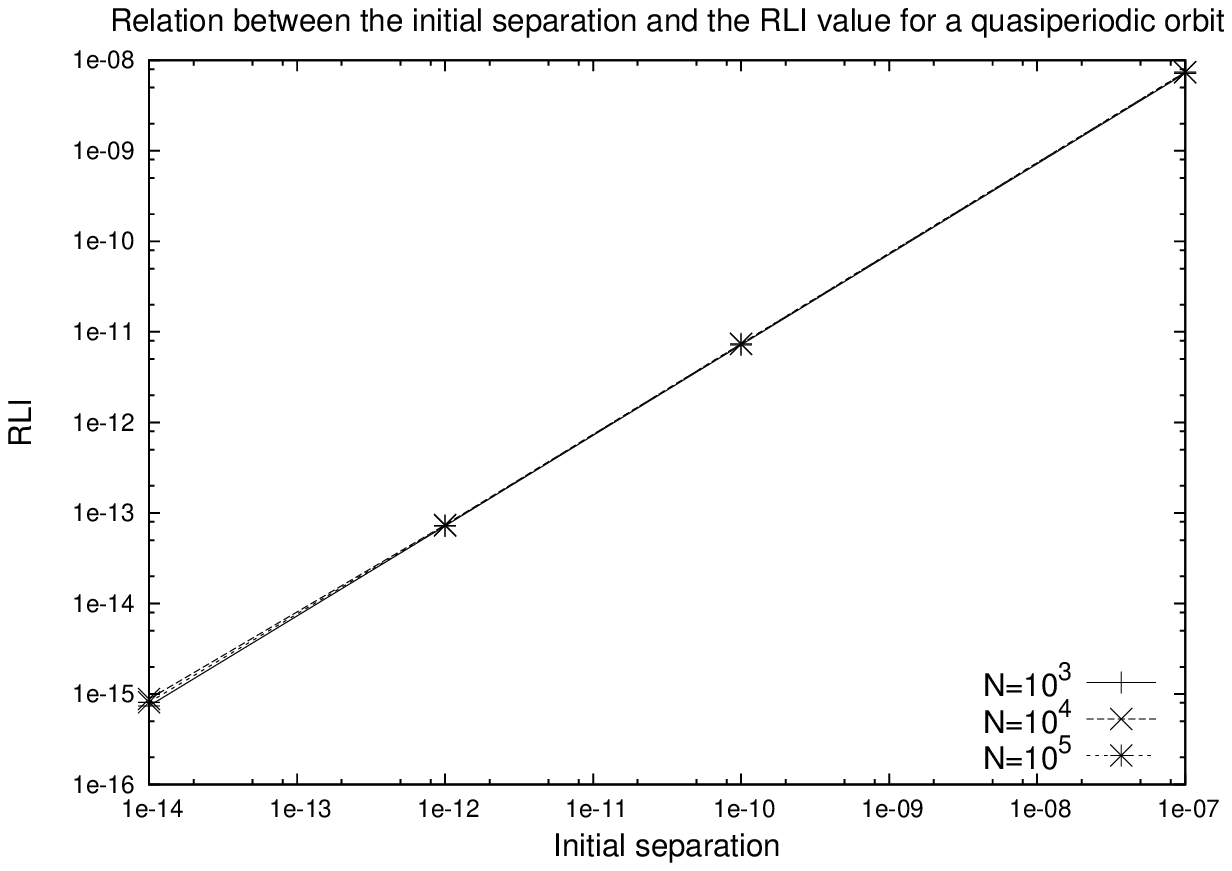}}& 
\hspace{-5mm}\resizebox{63mm}{!}{\includegraphics{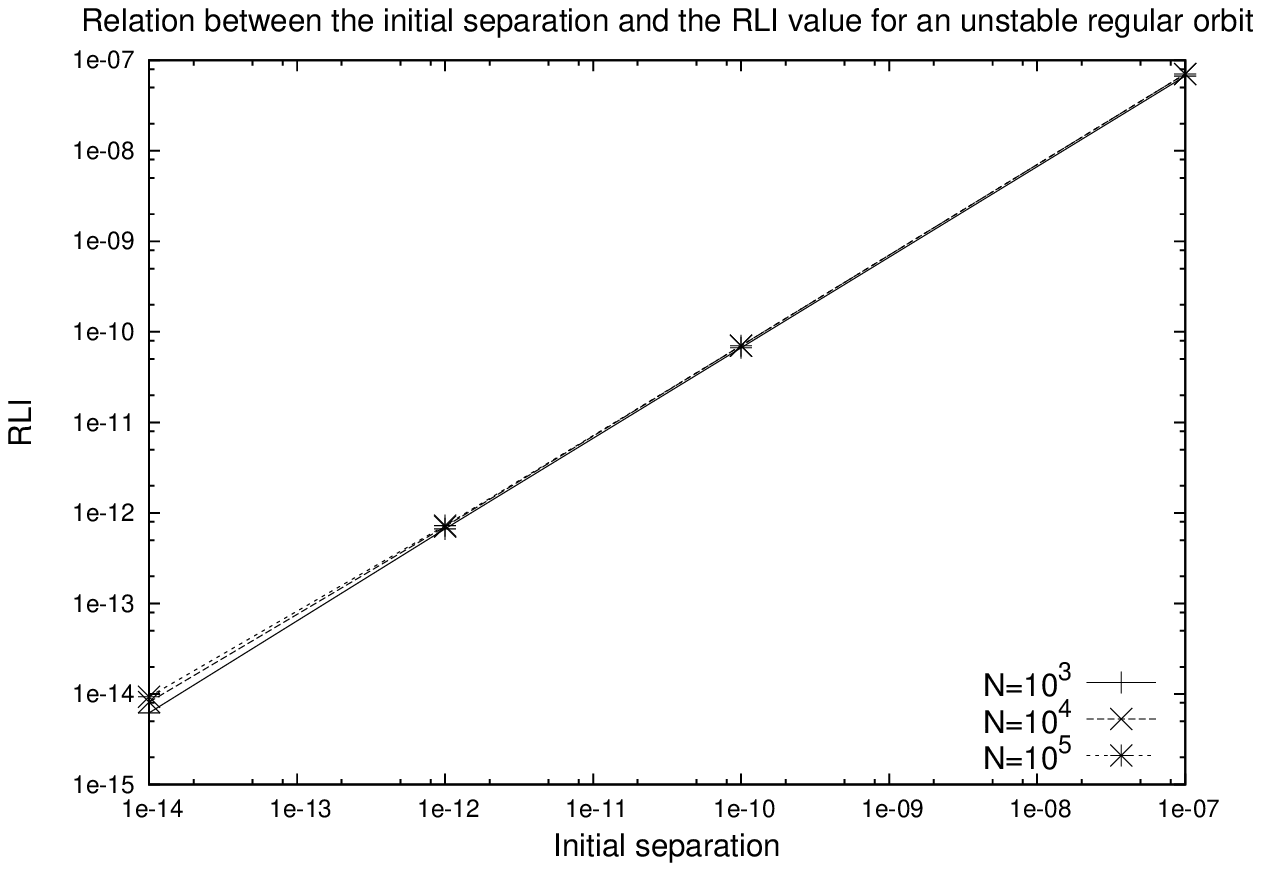}}
\end{tabular}
\caption{RLI values as a function of the initial separation for a pair of regular orbits and different $N$s. On the left, a quasiperiodic orbit, and on the right a regular orbit close to a hyperbolic object.} 
\label{rliinisep}
\end{center}
\end{figure}

In Fig. \ref{rliinisep}, we present the variation of the RLI final values for two regular orbits, a quasiperiodic orbit on the left, and a regular orbit close to a hyperbolic object on the right, and for four different values of $N$.
 
In both cases, the $N$ influences slightly the slope of the curves if the initial separation is small, i.e., below the value $10^{-12}$. However, if a threshold has to be determined, the proximity to a hyperbolic object affects the indicator in an order of magnitude (compare both Figs. \ref{rliinisep}). That is, if we decide to start the computation with an initial separation of $10^{-12}$, the relation shown in the left panel of Fig. \ref{rliinisep} tells us that a good estimate for the threshold is $10^{-13}$. Then, every orbit with a RLI final value above $10^{-13}$ would be classified as chaotic. On the other hand, if we consider the relation shown for the regular orbit with a certain level of instability (right panel of Fig. \ref{rliinisep}), the associated threshold for an initial separation of $10^{-12}$ is $10^{-12}$. That is, only those orbits with RLI final values above $10^{-12}$ will be considered as chaotic. Thus, taking the first threshold, $10^{-13}$, might lead to a misclassification of the nature of ''unstable'' regular orbits. 

The linear relationship between the threshold for the RLI and the initial separation parameter within the interval suggested by S\'andor et al. (2004) should be done by computing the RLI value for an orbit (or group of orbits) known, a priori, to be  regular but close to a hyperbolic object. 

In our case, the threshold selected and used in the previous experiments was $10^{-12}$ and the performance of the indicator was rather good. Nevertheless, in Section \ref{Qualitative study-representative groups}, top left panel of Fig. \ref{cao.reg.cvalue}, we found a regular orbit above the threshold. This can be explained by its great proximity to a stochastic layer. The higher the accuracy level required, the more careful the selection of the initial separation parameter should be; therefore, an iterative process might be advisable. 

\section{Discussion}\label{Discussion}
As every method has advantages and drawbacks, it is advisable to use different methods. Nevertheless, the aim of this work is finding, if possible, a ``CIs' function'' (hereafter, CIsF) which means a function of the CIs that represents the most efficient way to gather dynamical information from a mapping. Thus, we summarise here the procedure to study the phase space portrait of the vFSM using the most appropriate methods and explain the reasons of our choices. 

In order to employ the good performances shown by the RLI for large arrays of orbits and large $N$, we first need to calibrate the relationship between the initial separation parameter and its threshold. It is important to have different kinds of motion, from quasiperiodic to chaotic orbits. To have a good number of regular orbits with some amount of hyperbolicity is strongly advisable because they are the most influential type of orbits when fitting such relation (see Section \ref{A case of mild chaos}). Consequently, we need their location. Then, a fast overview of the whole phase portrait of the system is desirable and an expeditious CI with a theoretical threshold (which plays the role of an accurate guideline; an empirical threshold might not be so accurate) seems to be the best choice. According to Section \ref{Thresholds study}, the FLI has a theoretical time-dependent reliable threshold which seems to serve efficiently for a quick survey of the phase space portrait (see also the remarks about the speed of convergence in Section \ref{Qualitative study-statistical sample}, where the FLI shows very good results with a small $N$, due mainly to the versatility of its time-dependent threshold). Using the theoretical threshold of the FLI is advisable as a first attempt to locate individual orbits which will help to calibrate the empirical threshold of the RLI, as such CI has a better perfomance for big samples of orbits and larger values of $N$. 

After a quick glance at the phase space portrait with the FLI using a few iterations, it is possible to select some test orbits from different scenarios to calibrate the empirical threshold of the RLI. We chose some orbits close to the stochastic layer inside the main resonance (to select some regular orbits with an amount of hyperbolicity), and in the sticky region surrounding such resonance (in order to fix the parameters to distinguish them from the rest of the chaotic orbits and improve the description of the chaotic component). The preferred CIs for single studies are the MEGNO(2,0) and the SALI. Both CIs have different ways to identify chaotic or regular motion (see Section \ref{Qualitative study-representative groups}) and make the process of distinguishing the motions easier and faster. Besides, both CIs provided good results while revealing the true nature of sticky chaotic orbits. The $N_{sat}$ for the SALI is fundamental for the task (see Section \ref{Qualitative study-statistical sample} and Section \ref{Extreme conditions}). 

Finally, with this test orbits analyzed and an initial separation parameter for the RLI of $10^{-12}$ (which is a reliable choice for most dynamical situations, see S\'andor et al., 2004), the calibration of the RLI can be done (see Section \ref{A case of mild chaos}). A first good approach for the threshold of the RLI turns out to be $10^{-12}$ (an iterative process can continue until the desired level of accuracy). Therefore, applying the RLI to larger $N$ provides a more accurate global overview of the phase space portrait. The RLI seems to be the best choice because of its robust threshold and a well-suited power of resolution when dealing with big samples of orbits (see Section \ref{Thresholds study} and Section \ref{Qualitative study-statistical sample}). It has not a level of saturation for chaotic orbits, so the hyperbolicity levels are rather preserved independently of the $N$ (this can be better accomplished by the $N_{sat}$ of the FLI or the SALI). And the level of description for the regular component with the final values is higher than that obtained with the other methods revisited in this work (see Section \ref{Qualitative study-statistical sample}).

Summing up, the CIsF for the vFSM (and probably a good first approach for any mapping) is made up of:

1- the FLI final values (with the concomitant $N_{sat}$) to quickly identify the regions where test orbits can be selected in order to calibrate the other methods; 

2- the MEGNO(2,0) and the SALI (with the corresponding $N_{sat}$) are appropriate to analyze the test orbits or further interesting cases through the time evolution curves; 

3- the RLI final values to study globally the phase space portrait of the system on more stable regimes.

Two clear restrictions for the preceeding CIsF are: the iterative nature of the mappings and the number of indicators considered. Therefore, this work is nowadays being extended to deal with Hamiltonian flows and to encompass in the comparison not only the indicators tested so far, i.e.: the LI, the RLI, the MEGNO, the \textit{D}, the SSN, the SALI, and the FLI, but the GALI and the OFLI as well. All these CIs are being included in a code from which the user can select the appropriate CIsF from a variety of CIs. 

The introduction of both the APLE and the OFLI$^2_{TT}$ to complete the group of variational indicators will be considered in a future paper. Also, the FMA -an example of spectral indicator- will be tested against variational indicators in order to complete the comparison of CIs.

As a preliminary result of the investigation currently under way, we may state that a combination of the OFLI and the GALI seems to improve the choice of the FLI, the SALI and the RLI in the CIsF. 
 
\section*{Acknowledgments}
The authors want to thank both anonymous referees and the associate editor for their patience and dedication to improve the quality of the reported research. This work was supported with grants from the Consejo Nacional de Investigaciones Cient\'{\i}ficas y T\'ecnicas de la Rep\'ublica Argentina (CCT--La Plata) and the Universidad Nacional de La Plata.

\end{document}